\documentclass[12pt]{article}
\usepackage{amsfonts,amsthm,amsmath,amssymb,upgreek,bm}
\usepackage{hyperref}
\usepackage[paper=letterpaper,margin=.85in]{geometry}
\usepackage{graphicx}
\usepackage{units}
\usepackage{float}
\usepackage[export]{adjustbox}
\usepackage{xcolor}
\usepackage[shortlabels]{enumitem}
\usepackage[mathscr]{euscript}
\usepackage[normalem]{ulem}
\usepackage{bm}
\usepackage{tikz}
\usetikzlibrary{shapes,snakes}

\def\bpm{\begin{pmatrix}}
\def\epm{\end{pmatrix}}

\newcommand{\be}{\begin{equation}}
\newcommand{\ee}{\end{equation}}
\newcommand{\bea}{\begin{eqnarray}}
\newcommand{\eea}{\end{eqnarray}}
\renewcommand{\tilde}{\widetilde}
\renewcommand{\i}{\mathrm{i}}
\renewcommand{\d}{\mathrm{d}}

\newcommand{\mSFF}{K_E(t)}

\newcommand{\Tr}{\text{Tr}}
\newcommand{\STr}{\text{STr}}

\numberwithin{equation}{section}

\def\tr{\text{Tr}}
\def\Tr{\text{Tr}}

\begin{document}
\thispagestyle{empty}

\vspace*{2.5cm}
\begin{center}

{\bf {\LARGE A convergent genus expansion for the plateau}}\\

\begin{center}

\vspace{1cm}

{\bf Phil Saad$^1$, Douglas Stanford$^2$, Zhenbin Yang$^2$, and Shunyu Yao$^2$}\\
 \bigskip \rm
  
\bigskip
$^1$\hspace{.05em}School of Natural Sciences,\\ Institute for Advanced Study, Princeton, NJ 08540

\bigskip 

$^2$\hspace{.05em}Stanford Institute for Theoretical Physics,\\Stanford University, Stanford, CA 94305
\rm
  \end{center}

\vspace{2.5cm}
{\bf Abstract}
\end{center}
\begin{quotation}
\noindent

We conjecture a formula for the spectral form factor of a double-scaled matrix integral in the limit of large time, large density of states, and fixed temperature. The formula has a genus expansion with a nonzero radius of convergence. To understand the origin of this series, we compare to the semiclassical theory of ``encounters'' in periodic orbits. In Jackiw-Teitelboim (JT) gravity, encounters correspond to portions of the moduli space integral that mutually cancel (in the orientable case) but individually grow at low energies. At genus one we show how the full moduli space integral resolves the low energy region and gives a finite nonzero answer.

\end{quotation}

\setcounter{page}{0}
\setcounter{tocdepth}{2}
\setcounter{footnote}{0}
\newpage

\parskip 0.1in
 
\setcounter{page}{2}
\tableofcontents

\newpage

\section{Introduction}

A longstanding challenge is to explain the discrete spectrum of black hole microstates using spacetime geometry. In recent years, some statistical aspects of these microstates have been explained using spacetime wormholes.\footnote{The role of spacetime wormholes in quantum gravity has also been a longstanding puzzle, \cite{Hawking:1987mz,Lavrelashvili:1987jg,Giddings:1987cg,Coleman:1988cy,Maldacena:2004rf,ArkaniHamed:2007js}.} Examples include: aspects of the spectral form factor \cite{Saad:2018bqo,Saad:2019lba,Stanford:2019vob} and late-time correlation functions \cite{Blommaert:2019hjr,Saad:2019pqd,Blommaert:2020seb,Stanford:2021bhl}, the Page curve \cite{Almheiri:2019qdq,Penington:2019kki} and matrix elements \cite{Stanford:2020wkf} of an evaporating black hole, and the ETH behavior of matrix elements \cite{Belin:2020hea,Belin:2021ryy,Chandra:2022bqq}. 

A statistical theory of microstates is far from a complete description, but it is enough to probe discreteness of the energy spectrum. One tool to discuss this is the spectral form factor
\be\label{intro:SFF}
K_\beta(t) = \langle Z(\beta+\i t)Z(\beta - \i t)\rangle
\ee
where $Z(x) = \tr\,e^{-x H}$ is the thermal partition function, and the brackets represent some form of averaging for which the statistical description is sufficient. The discrete nature of chaotic energy levels is reflected in the ``plateau'' to a late time value $K_\beta(\infty) = Z(2\beta)$. 

For systems in the unitary symmetry class (no time reversal symmetry), the random matrix theory (RMT) prediction for the spectral form factor is simple. The microcanonical version
\be
K_E(t) = \int \frac{\d\beta}{2\pi \i} e^{2\beta E} K_\beta(t)
\ee
should have the form  of a linear ramp connected to a plateau: $\text{min}\{t/2\pi, e^{S(E)}\}$ with $S(E)$ the microcanonical entropy at energy $E$. This sharp transition at $t_p = 2\pi e^{S(E)}$ is a signature of discretness of the spectrum; it arises from oscillations in the density pair correlator with wavelength $e^{-S(E)}$, representing the mean spacing between discrete energy levels.

The sharpness of the transition from the ramp to the plateau is an apparent obstruction to an explanation in terms of geometry. In particular, in two-dimensional dilaton gravity models such as JT gravity, the genus expansion should roughly be thought of as an expansion in $e^{-S(E)}$. But the transition from the ramp to the plateau comes from contributions have go as $e^{i \# e^{S(E)}}$, nonperturbative in the genus counting parameter, suggesting that it is not captured by the conventional sum over geometries.\footnote{Some previous approaches to explaining the plateau through a sum over geometries have involved ``spacetime D-branes" \cite{Saad:2019lba,Blommaert:2019wfy,Marolf:2020xie}, which generalize the sum over geometries to include contributions from an infinite number of asyptotic boundaries.}

However, the spectral form factor $K_\beta(t)$ is an integral of $K_E(t)$ over energy, and this integral has the potential to smooth out the transition to the plateau. As first shown by \cite{Okuyama:2020ncd,okounkov2002generating} for the Airy matrix integral, the resulting function can have a convergent genus expansion, smoothly transitioning from the ramp to the plateau. We conjecture a generalization of this result below, in a limit that will be referred to as ``$\tau$-scaling.'' This convergent series makes it possible to explain the plateau in terms of a conventional sum over geometries, rather than from a radical nonperturbative effect. 

In this paper, we will explain some features of this genus expansion for the spectral form factor, primarily working in the low-energy limit of JT gravity: the Airy model. Our explanations will connect with the encounter computations in semiclassical periodic orbit theory, used to explain the RMT corrections to the ramp \cite{2009arXiv0906.4930A,Sieber_2001,Sieber_2002,PhysRevLett.93.014103,PhysRevE.72.046207}. The sum over encounters is closely analogous to a genus expansion, so it is natural to try interpret the genus expansion for the plateau in terms of a gravitational analog of encounters. Encounters alone cannot be sufficient to explain the genus expansion for $K_\beta(t)$ because without time-reversal symmetry, the encounters cancel genus by genus.

The models that we study, in particular the Airy model, allow us to generalize the theory of encounters beyond their usual regime of validity in the high-energy, semiclassical limit. At very low energies, of order $1/t$, the encounters receive large quantum corrections that disturb the cancellation between encounters, reproducing the expected $\tau$-scaled $K_\beta(t)$.

In \hyperref[SectionTwo]{\textbf{Section Two}}, we introduce a formula for $K_\beta(t)$ in a double-scaled matrix integral in the ``$\tau$-scaled" limit, generalizing \cite{Okuyama:2020ncd,Okuyama:2021cub}. We reconcile the existence of a convergent genus expansion for $K_\beta(t)$ with the absence of such an expansion for $K_E(t)$. In particular, one can think of the genus expansion for $K_\beta(t)$ as coming entirely from very low energies.

In \hyperref[SectionThree]{\textbf{Section Three}} we review an analog of the genus expansion for $K_E(t)$ in periodic orbit theory: the sum over encounters. The sum over encounters gives an expansion in $e^{-S(E)}$, valid at high energies. For periodic orbit systems in the GUE symmetry class (no time-reversal), corrections to the ramp coming from encounters cancel order by order \cite{PhysRevLett.93.014103,PhysRevE.72.046207}. In JT gravity, we discuss a direct analog of the simplest type of encounter contribution in a theory with time-reversal symmetry, contributing to the SFF at genus one-half.

In \hyperref[SectionFour]{\textbf{Section Four}} we study the Airy model, the low-energy limit of JT gravity. The wormhole geometries in this model are very simple, and in one-to-one correspondence with ribbon graphs in the Feynman diagram expansion of Kontsevich's matrix model. These graphs allow us to generalize the encounter computations beyond the semiclassical, high-energy regime. At genus one and high energies, the encounter contributions mutually cancel in the GUE symmetry class. At low energies, quantum corrections to the encounters spoil this cancellation, leading to the nonzero contribution to $K_\beta(t)$. The full answer at this genus comes from a large region of moduli space, far from the semiclassical encounter regime.

{\bf Note:} Two recent papers \cite{Blommaert:2022lbh,Weber:2022sov} are closely related to our work. A preliminary version of section two of this paper was shared with the authors of \cite{Blommaert:2022lbh,Weber:2022sov} in October 2021.

\section{Tau scaling of the spectral form factor}\label{SectionTwo}
In this section we discuss the ``$\tau$-scaling'' limit of matrix integrals in which we conjecture that the spectral form factor has a simple form with a convergent genus expansion. Consider a double-scaled matrix integral with unitary symmetry class and classical density of states
\be
\rho(E) = e^{S_0}\rho_0(E).
\ee
The spectral form factor is defined as
\be
K_\beta(t) \equiv \langle Z(\beta+\i t)Z(\beta - \i t)\rangle, \hspace{20pt} Z(x) \equiv \tr\, e^{-x H}.
\ee
Here the angle brackets represent the average in the matrix integral. We would like to analyze this in a limit where $t$ goes to infinity and $e^{S_0}$ also goes to infinity, holding fixed $\beta$, and also holding fixed the ratio
\be
\tau = t e^{-S_0}.
\ee
This will be referred to as the ``$\tau$-scaled'' limit. 

In the $\tau$-scaled limit, the time $t = e^{S_0}\tau$ is large, so the SFF will be dominated by correlations of nearby energy levels. Pair correlations of nearby levels are described by the universal sine-kernel formula, which translates to a ramp-plateau structure $\text{min}\{t/2\pi,\rho(E)\}$ as a function of the center of mass energy $E$. By integrating this contribution over $E$, one gets the following candidate expression for the spectral form factor
\begin{align}\label{foldingformula}
K_\beta(t) &\stackrel{?}{\approx} \int_{E_0}^\infty \d E \ e^{-2\beta E} \text{min}\left\{\frac{t}{2\pi},\rho(E)\right\}.
\end{align}
This was previously discussed as an uncontrolled approximation to the SFF \cite{Cotler:2016fpe}. Here we would like to propose that it is exact in the $\tau$-scaled limit, 
\be
\lim_{S_0\to\infty} e^{-S_0}K_\beta(\tau e^{S_0}) = \int_{E_0}^\infty \d E \ e^{-2\beta E}\text{min}\left\{\frac{\tau}{2\pi},\rho_0(E)\right\}.\label{tauconj}
\ee

Let's try an example by taking $\rho_0(E) = \frac{\sqrt{E}}{2\pi}$, which is sometimes called the Airy model, or the Kontsevich-Witten model. Then (\ref{tauconj}) becomes
\begin{align}
e^{-S_0}K_\beta(\tau e^{S_0}) &= \frac{1}{2\pi}\int_0^\infty \d E e^{-2\beta E}\text{min}\left\{\tau,\sqrt{E}\right\}\\ 
&= \frac{1}{2\pi}\frac{\pi^{1/2}}{2^{5/2}\beta^{3/2}}\text{Erf}(\sqrt{2\beta}\tau)\label{airyconj}\\
&=\frac{\tau}{4\pi\beta} - \frac{\tau^3}{6\pi}+ \frac{\beta}{10\pi}\tau^5 - \frac{\beta^2}{21\pi} \tau^7 +\dots.\label{airyanssec1}
\end{align}
We can compare this to the exact answer for the spectral form factor of the Airy model \cite{okounkov2002generating,Okuyama:2021cub}
\begin{align}
K_\beta(t) &=  \langle Z(2\beta)\rangle \text{Erf}(e^{-S_0}\sqrt{2\beta(\beta^2+t^2)})\\
&=  \frac{\exp\left(S_0 + \frac{1}{3}e^{-2S_0}\beta^3\right)}{4\sqrt{\pi}\beta^{3/2}}\text{Erf}(e^{-S_0}\sqrt{2\beta(\beta^2+t^2)}).\label{eqn:airySFF}
\end{align}
This agrees with (\ref{airyconj}) in the $\tau$-scaled limit. 

As a second example, we can take $\rho_0(E) =\frac{1}{4\pi^2}\sinh(2\pi\sqrt{E})$, which corresponds to JT gravity:
\begin{align}
e^{-S_0}K_\beta(\tau e^{S_0}) &= \frac{1}{2\pi} \int_0^\infty \d E e^{-2\beta E} \text{min}\left\{\tau,\frac{1}{2\pi}\sinh(2\pi\sqrt{E})\right\}\\
&= \frac{e^{\frac{\pi^2}{2\beta}}}{16\sqrt{2\pi}\beta^{3/2}}\left[\text{Erf}\left(\frac{\frac{\beta}{\pi}\text{arcsinh}(2\pi \tau) + \pi}{\sqrt{2\beta}}\right)+\text{Erf}\left(\frac{\frac{\beta}{\pi}\text{arcsinh}(2\pi \tau)- \pi}{\sqrt{2\beta}}\right)\right]\label{JTerfs}\\
&= \frac{\tau}{4\pi\beta} - \frac{\tau^3}{6\pi}+ \left(\frac{\beta}{10\pi}+\frac{2\pi}{15}\right)\tau^5 - \left(\frac{\beta^2}{21\pi} + \frac{4\pi\beta}{21} + \frac{64\pi^3}{315}\right)\tau^7 +\dots\label{tauexp}
\end{align}
For JT gravity, no exact formula for the SFF is known,\footnote{See \cite{Okuyama:2020ncd,Okuyama:2021cub} for discussion of a different limit where $\beta$ is also large, and see \cite{Johnson:2020exp} for numerical evaluation.} but (\ref{JTerfs}) can be checked by using topological recursion \cite{Eynard:2004mh,Eynard:2007kz} to compute the exact spectral form factor to a given order in $e^{-S_0}$, and then applying $\tau$-scaling. Using this method, we confirmed (\ref{JTerfs}) up to order $\tau^{13}$. 

Note that the series in $\tau$ corresponds to the genus expansion, as one can see by undoing the $\tau$-scaling and replacing $\tau \to t e^{-S_0}$. In particular, the power of $\tau$ is $\tau^{2g+1}$. Normally, the genus expansion of the SFF in JT gravity is an asymptotic series. But after $\tau$-scaling it has a nonzero radius of convergence $|\tau| < {1\over 2\pi}$, and the analytic continuation is nonsingular along the entire real $\tau$ axis. For large $\tau$ it reproduces the plateau.\footnote{In appendix \ref{Pappendix} we show that (\ref{tauconj}) always has a nonzero radius of convergence.}

The presence of powers of $\beta$ in the leading $\tau$-scaled answer indicates that there are cancellations of higher powers of $t$. For example, the term proportional to $\beta \tau^5$ arises from a linear combination of terms $(\beta + \i t)^{p_1}(\beta - \i t)^{p_2}$ with $p_1 + p_2 = 6$ such that the leading power $t^6$ cancels. This cancellation has been studied by \cite{Blommaert:2022lbh,Weber:2022sov}.

The conjecture (\ref{tauconj}) was designed so that if we compute the inverse Laplace transform to $K_E(\tau e^{S_0})$, the answer will simply be $e^{S_0}\min\{\tau/2\pi,\rho_0(E)\}$. In particular, for fixed $E > 0$, the expansion in powers of $\tau$ terminates after the linear term -- naively there is simply no genus expansion for fixed energy in the $\tau$-scaled limit. A more refined viewpoint is that the genus expansion has coefficients that are derivatives of $\delta$ functions of $\rho(E)$. This can be seen by writing $\min(x,y) = x - (x-y)\theta(x-y)$ and expanding in powers of $x$. It can also be seen by inverse Laplace transforming  (\ref{tauexp}) term by term. 

So the genus expansion of the canonical SFF can be understood as arising from contributions localized at zero energy where the plateau time is short. To see this from another perspective, consider a $\rho_0$ of the general form
\be\label{generaldensity}
2\pi\rho_0(E) = a_1 E^{1/2} + a_3 E^{3/2} + a_5 E^{5/2} + a_7 E^{7/2}+\dots
\ee
Then the conjecture (\ref{tauconj}) gives 
\begin{align}\label{tauexpansionintro}
\int \d E e^{-2\beta E} &\text{min}\left\{\frac{\tau}{2\pi},\rho_0(E)\right\}\\& = \frac{\tau}{4\pi\beta} - \frac{\tau^3}{6\pi a_1^2} + \frac{(a_1\beta + 2a_3)}{10 \pi a_1^5}\tau^5 - \frac{(2a_1^2\beta^2 + 12 a_1a_3\beta - 6 a_1 a_5 + 21 a_3^2)}{42\pi a_1^8}\tau^7+\dots\notag
\end{align}
The contribution from genus $g$ depends on only the $g-1$ first terms in the expansion of $\rho_0(E)$ around $E = 0$. Indeed, in appendix \ref{Pappendix} we show that the coefficient of $\tau^{2g+1}$ for $g \ge 1$ is
\be
-\frac{1}{g(2g+1)(2\pi)^{2g+1}} \oint_0 \frac{\d E}{2\pi \i} \frac{e^{-2\beta E}}{\rho_0(E)^{2g}}.
\ee

In the rest of the paper we will try to understand where this series comes from. We will start by comparing to another type of expansion associated to the spectral form factor -- the theory of encounters in periodic orbits.

\newpage

\section{Encounters in orbits and in JT}\label{SectionThree}
One case where the spectral form factor has been studied extensively is semiclassical chaotic billiards.\footnote{See the introduction of \cite{muller2005periodic} for history and references.} There, the Gutzwiller trace formula is used to write an expression for the spectral form factor in terms of a sum over pairs of periodic orbits. Special pairings of orbits called ``encounters'' lead to a series in $\tau$ that is vaguely reminiscent of (\ref{tauexpansionintro}).

However, there are important differences: the encounters cancel between themselves for systems with unitary symmetry class, and the encounter analysis is only valid at high energies, in the semiclassical region. It is tempting to view the genus expansion (\ref{tauexpansionintro})  as analogous to a type of ``souped up'' encounter theory that can accurately treat very low energies, outside the semiclassical limit, and for which the encounters do not quite cancel.

We will explore this further in section \ref{SectionFour}. In the current section we prepare by reviewing the theory of encounters in periodic orbits and finding an analog of the simplest (Sieber-Richter) encounter in a JT gravity calculation.

\subsection{Review of periodic-orbit theory}
Consider a semiclassical billiards system, consisting of a particle moving in a stadium.\footnote{We set $\hbar = 1$. The semiclassical limit corresponds to high energies.} The starting point for the theory of encounters is Gutzwiller's trace formula for the oscillating part of the density of states $\rho_{\text{osc}}(E)$ in terms of a sum over classical periodic orbits $\gamma$:
\be
\rho_{\text {osc}}(E)\sim {1\over \pi }\text{Re}\sum_{\gamma} A_{\gamma} e^{\i S_{\gamma}}.
\ee
Here $A_{\gamma}$ is the stability amplitude (one-loop determinant) and $S_{\gamma}$ is the classical action. The microcanonical spectral form factor is then given by a double sum over orbits $\gamma,\gamma'$:
\begin{align}
\mSFF &=\langle \int \d\epsilon e^{\i \epsilon t} \rho_{\text{osc}}(E+{\epsilon \over 2})\rho_{\text{osc}}(E-{\epsilon\over 2})\rangle \\
&={1\over 2\pi }\langle \sum_{\gamma,\gamma'} A_{\gamma}A^*_{\gamma'}e^{\i (S_{\gamma}-S_{\gamma'})}\delta\left(t-{t_{\gamma}+t_{\gamma'}\over 2}\right)\rangle.\label{eqn:SFF}
\end{align}
Here $t_{\gamma}={\partial S_{\gamma}\over \partial E}$ is the period of the semiclassical orbits $\gamma$ and $\langle \cdot \rangle$ represents an average over the energy window.

$K_E(t)$ receives both diagonal ($S_\gamma=S_\gamma'$) and off-diagonal ($S_\gamma\neq S_\gamma'$) contributions. In a chaotic system, one expects $S_{\gamma}=S_{\gamma'}$ only if $\gamma$ and $\gamma'$ are identical or related by symmetry -- the simplest (GUE) case is to assume there is no symmetry so $\gamma = \gamma'$. Berry showed \cite{berry1985semiclassical} that the sum over $\gamma = \gamma'$ leads to the linear ramp  $t/2\pi$ in the GUE spectral form factor. The factor of $t$ comes from the possibility of a relative time shift between $\gamma$ and $\gamma'$. In the GOE case, there is additional time reversal symmetry $\mathcal{T}^2=1$, and diagonal sum also contains the time reversed orbit $\gamma'=\mathcal{T}\gamma$. This leads to an additional factor of two, so $K_E(t) \sim t/\pi$.

The off-diagonal contributions are weighted by an oscillatory factor $e^{\i (S_{\gamma}-S_{\gamma'})}$. Encounter theory is a way of identifying systematic classes of orbits such that the difference in actions is small. These consist of orbit pairs $\gamma,\gamma'$ that closely follow each other except for small off-shell regions known as encounters. The impressive achievement of encounter theory is that a sum over such encounters reproduces the fact that the GUE $K_E(t)$ has no corrections before the plateau, and the GOE $K_E(t)$ has a particular expansion
	\begin{align}
		K_E^{(GOE)}(t)&={t\over \pi}-{t\over 2\pi}\log\left[1+{t\over \pi  \rho(E)}\right]\\ &={t\over \pi}-{2t^2\over\rho(E)(2\pi)^2}+{2 t^3\over \rho(E)^2(2\pi)^3}-...
	\end{align}
Berry's analysis explains the linear term. The quadratic term was explained by Sieber and Richter \cite{Sieber_2001}, the cubic term was explained in \cite{heusler2004universal}, and the full series was reproduced in \cite{muller2005periodic}.

\subsubsection{Sieber-Richter pair}
The simplest example of an encounter is the Sieber-Richter pair or ``2-encounter'' which exists in a theory with time-reversal symmetry. The pair of orbits $\gamma,\gamma'$ can be sketched in configuration space as follows (this figure and (\ref{fig223}) were modified from \cite{muller2005periodic} with permission):
\be\label{SRfig}
\includegraphics[valign = c, scale = 0.4]{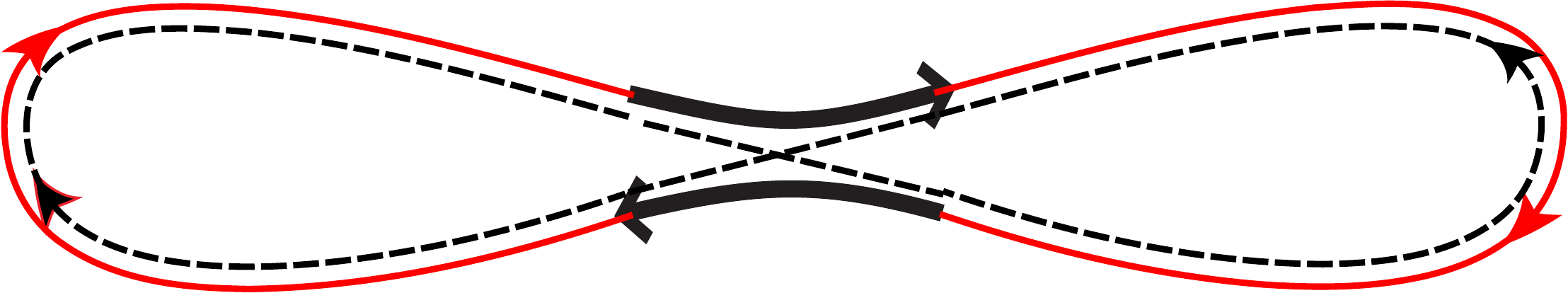}
\ee
We will focus on the case with only two degrees of freedom. 
The key feature is that the orbit $\gamma$ (the red/solid orbit) returns close to itself at some point $t_1$ along the orbit. This point is referred to as an encounter, and the partner orbit $\gamma'$ differs from $\gamma$ only in the vicinity of the encounter (and as a consequence it is time-reversed in one of the two ``stretches'' outside the encounter region).  

The encounter can be characterized by the deviation of the two nearby segments of $\gamma$, and it is convenient to decompose this deviation into the stable and unstable directions $s,u$. Within the encounter region, the $s,u$ variables decay and grow exponentially in time, with a Lyapunov exponent $\lambda$. This determines the duration of the encounter region:
 \be
 t_{enc}={1\over \lambda}\log {c^2\over |s u|},
 \ee
where $c$ characterizes the regime of validity of the linearized analysis near the encounter.\footnote{At high energies, the result does not depend on the precise value of $c$.} In the two regions outside the encounter (called stretches), the two orbits follow each other closely, up to time reversal. This means that the difference in actions $S_\gamma  - S_{\gamma'}$ comes only from the encounter region itself. This difference in action is determined by the $s,u$ variables and takes the form\footnote{This is reminiscent of the action that controls out-of-time-order correlators \cite{Stanford:2021bhl,Gu:2021xaj}.}
\begin{equation}
	S_{\gamma}-S_{\gamma'}=s u.
\end{equation}

The probability that orbit $\gamma$ will have such an encounter is determined by ergodicity, which gives a uniform measure in the phase space ${1\over (2\pi)^2\rho_E}\d t_1 \d s \d u$. The Sieber-Richter pair's contribution to the spectral form factor can then be evaluated using the following integral:
\begin{equation}\label{srint}
	K_{E}(t)\supset {t\over \pi }{1\over  (2\pi)^2\rho(E)}\int^{c}_{-c} \d s \int_{-c}^{c}\d u{t\over 2t_{enc}} \int_{0}^{t-2t_{enc}} \d t_1 e^{\i su}.
\end{equation}
The factors in the integral are explained as follows:
\begin{enumerate}
	\item The overall ${t\over \pi }=2\times {t\over 2\pi }$ factor reflects the relative time shift between $\gamma$ and $\gamma'$ and the time reversal symmetry. This part is the same as in the linear ramp.
	\item The additional ${t\over 2t_{enc}}$ factor reflects the fact that the encounter region can be anywhere along the orbit: $t$ comes from integrating over the time of the reference point; ${1\over 2t_{enc}}$ fixes an over-counting from the choice of the reference point inside the encounter (changing this reference point would rescale $s$ and $u$ oppositely).

	\item The integration range of the time where the encounter takes place, $t_1$,  is upper bounded by $t-2t_{enc}$ to ensure the existence of the encounter region.
\end{enumerate}   
The integral (\ref{srint}) gives:
\begin{equation}\label{eqn:k1/2}
	K_E(t)\supset {t\over \pi }{t\over  (2\pi)^2\rho(E)}\int \d s \d u e^{\i su}({t\over 2t_{enc}}-1)\approx-{2t^2\over (2\pi)^2\rho(E)}.
\end{equation}
Naively the answer should be of order $t^3$, but this term is proportional to $\int \d s \d u e^{\i s u}/\log|su| \approx 0$, and the nonzero answer comes from the subleading $t^2$ term.

\subsubsection{Cancellation of encounters in GUE}
The Sieber-Richter pair does not contribute in a theory without time-reversal symmetry (GUE case) because the portions of the orbits in the right stretch of figure (\ref{SRfig}) would have no reason to follow each other. Instead, in the GUE case the leading encounters (in the ${1\over \rho_E}$ expansion) are a configuration with two 2-encounters, and a configuration with a single 3-encounter where three segments of the orbit simultaneously approach each other:
\be\label{fig223}
\includegraphics[valign = c, scale = .382]{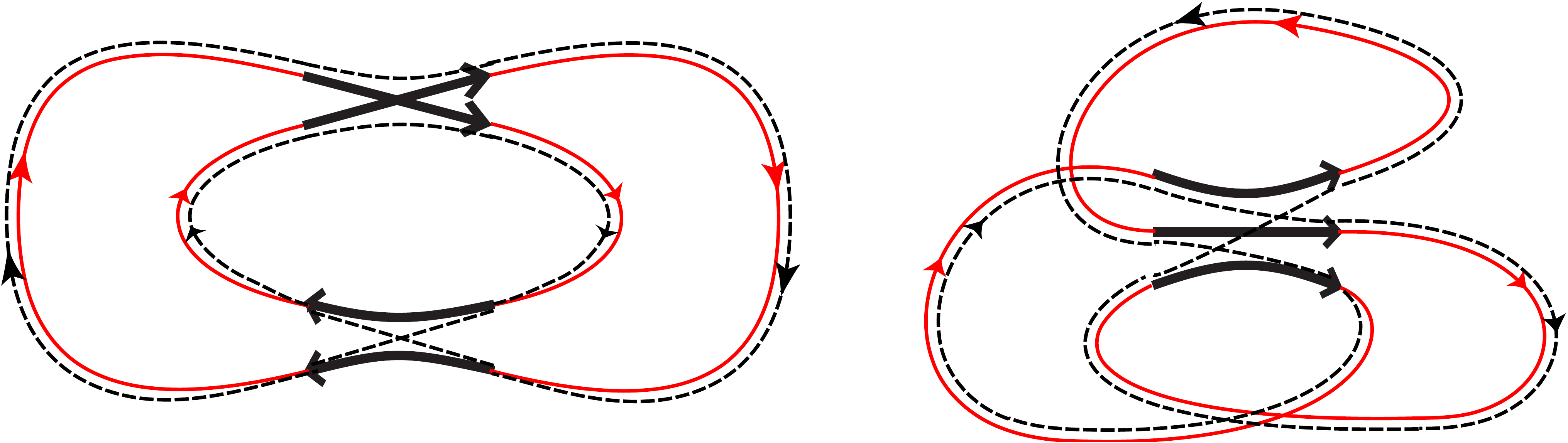}
\ee

The two 2-encounters (denoted as $(2)^2$) is a straightforward generalization of the 2-encounter.  It contains two pairs of $(s_i,u_i)$ soft modes with encounter time $t_{enc}^i$ and three zero modes $t_i$ labelling the stretch lengths.
Its contribution $K_{E,(2)^2}(t)$ to the spectral form factor is given by the following integral :
\be
K_E(t)\supset K_{E,(2)^2}(t)= {t\over 2\pi}{1\over (4\pi^2\rho(E))^2}\int \prod_{i=1}^2\d s_i \d u_i e^{\i \sum_{i=1}^2s_i u_i} {t\over 4 \prod_{i=1}^2t_{enc}^i} {(t-\sum_{i=1}^22t_{enc}^i)^3\over 6},
\ee
where the ${(t-\sum_{i=1}^22t_{enc}^i)^3\over 6}$ comes from the integration of the three zero modes $t_i$. As before, the $s_i,u_i$ integral is nonzero only when the measure is independent of $t_{enc}^i$. 
This kills the $t^5,t^4$ powers in the two 2-encounter and left with only a $t^3$ piece:
\be\label{twoEncounters}
K_{E,(2)^2}(t)={t^3\over (2\pi)^3\rho(E)^2}.
\ee 

The 3-encounter (denoted as $(3)^1$) is a limiting case of the two 2-encounters where one of the stretches shrinks to zero.
It can be thought of as a sequential swap of pairs of trajectories where each swap leads to an action difference $s_i u_i$ between the swapped trajectories. These deviations $(s_i,u_i)$ can be relatd to the deviations between nearest neighbor trajectories $(\hat s_i, \hat u_i)$\footnote{See Sec.II of \cite{muller2005periodic} for a detailed discussion.}, which determine the encounter duration $t_{enc}={1\over \lambda }\log{c^2\over \text{max}(\hat s_i)\text{max}(\hat u_i)}$. The contribution to the spectral form factor is
\begin{align}
K(t)\supset K_{E,(3)^1}(t)&={t\over 2\pi}{1\over (4\pi^2\rho(E))^2}\int \prod_{i=1}^2\d s_i \d u_i e^{\i \sum_{i=1}^2s_i u_i}  {t\over 3 t_{enc}} {(t-3t_{enc})^2\over 2}\\ &=-{t^3\over (2\pi)^3\rho(E)^2}.\label{threeencounter}
\end{align}
In particular, these two contributions cancel (although GOE variants of them that include the possibility of time-reversed stretches do not cancel). In \cite{muller2005periodic} it was shown that this cancellation between the GUE encounters continues to hold to all orders in the ${1\over \rho_E}$ expansion, reproducing the RMT expectation that the ramp is exact before the plateau time.

\subsection{Sieber-Richter pair in JT}\label{section:SRJT}
In this section, we will explain the analog of Sieber-Richter pair in JT gravity. As explained in \cite{Altland:2020ccq}, this corresponds to the topology of a cylinder with a crosscap inserted. We are grateful to Adel Rahman for collaboration on the calculations in this section.

The cylinder with a crosscap inserted corresponds to the following quotient of hyperbolic space:
\be
\includegraphics[valign = c, scale = 1.1]{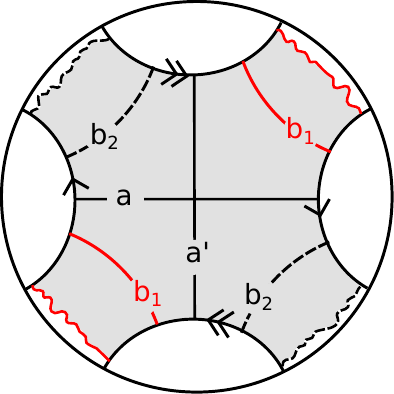}\label{embeddingdiagram}
\ee
The identification that defines the quotient is specified by gluing together the two geodesics with single arrows and also gluing together the two geodesics with double arrows, keeping mind the orientation of the arrows. 

The wiggly solid red segments form a single $S^1$ boundary, and have renormalized length $\beta_1$, which will be continued to $\beta + \i t$. Similarly, the wiggly dashed black segments form a single $S^1$ of renormalized length $\beta_2$, which will be continued to $\beta - \i t$. 

The two curves labeled $b_1$ together form a circular geodesic of length $b_1$, and the two curves labeled $b_2$ form a circular geodesic of length $b_2$. The two lines labeled $a,a'$ form two circular geodesics that intersect at a point. These are ``one-sided'' geodesics, meaning that a neighborhood of either one is a Mobius strip, rather than a cylinder. Hyperbolic geometry imposes one constaint on these parameters \cite{Norbury}:
\be\label{CCrelation}
\sinh(\frac{a}{4})\sinh(\frac{a'}{4})=\cosh(\frac{b_1+b_2}{4})\cosh(\frac{b_1-b_2}{4}).
\ee

The geometry has a $\mathbb{Z}_2$ mapping class group that interchanges $a\leftrightarrow a'$. A convenient way to fix this is to parametrize the geometry by $a$ and to require that $a < a'$. This amounts to requiring $a < a_*$, where
\be\label{CCcutoff}
\sinh^2(\frac{a^*}{4})=\cosh(\frac{b_1+b_2}{4})\cosh(\frac{b_1-b_2}{4}).
\ee

The path integral of JT gravity on this space is then
\be
2\times e^{-S_0}\int_0^\infty b_1\d b_1 b_2\d b_2 Z_{\text{Tr}}(\beta_1,b_1)Z_{\text{Tr}}(\beta_2,b_2)\int_{\epsilon}^{a^*} \frac{\d a}{2\tanh(\frac{a}{4})}.
\ee
Let's explain each of the factors in this expression. The factor of two is from the possibility of an orientation reversal on going from one boundary to the other. The factor of $e^{-S_0}$ is from the topological weighting $e^{S_0\chi}$ where $\chi = -1$ is the Euler characteristic of the crosscap cylinder. The integral over $b_1$ comes with a factor of $b_1$ that represents the integral of the Weil-Petersson measure $\d b \wedge \d\tau$ over the twist parameter $\tau$. The factors of $Z_{\text{Tr}}(\beta,b)$ represent the integral over the boundary wiggles. Finally, the $a$ parameter is integrated with the crosscap measure \cite{Norbury,Gendulphe,Stanford:2019vob} with an upper limit specified by $a_*$ to account for the mapping class group. Note that the integral would be divergent near $a = 0$, which represents the fact that in JT gravity, the path integral on non-orientable surfaces is divergent. We regularized the integral by cutting it off at $\epsilon$, and we will see that this divergence does not survive the $\tau$-scaling limit.\footnote{This regularization corresponds e.g.~to studying the $(2,p)$ minimal string with large but finite $p$.}

To obtain the contribution to the spectral form factor, we continue the parameters $\beta_1,\beta_2$ to $\beta \pm \i t$:
\be\label{canonicalCC}
\begin{aligned}
K_{\beta}(t)&\supset 2\times e^{-S_0}\int_0^\infty b_1\d b_1 b_2\d b_2\int_{\epsilon}^{a^*} \frac{\d a}{2\tanh(\frac{a}{4})} Z_{\text{Tr}}(\beta+it,b_1)Z_{\text{Tr}}(\beta-it,b_2) \\
&= 2\times e^{-S_0}\int_0^\infty b_1\d b_1 b_2\d b_2\int_{\epsilon}^{a^*} \frac{\d a}{2\tanh(\frac{a}{4})} \frac{e^{-b_1^2/(4(\beta+\i t))}}{\sqrt{4\pi (\beta+\i t)}}\frac{e^{-b_2^2/(4(\beta-\i t))}}{\sqrt{4\pi (\beta-\i t)}} \\
&\approx \frac{e^{-S_0}}{2\pi t}\int_0^\infty b_1\d b_1 b_2\d b_2\int_{\epsilon}^{a^*} \frac{\d a}{2\tanh(\frac{a}{4})} \exp\left\{\i\frac{b_1^2-b_2^2}{4t} - \frac{\beta}{4t^2}(b_1^2+b_2^2)\right\}.
\end{aligned}
\ee
Here, the lower bound  $\epsilon$ of the integration range of $a$ is the cutoff that regularizes the crosscap integral mentioned before -- it will drop out below. In the last step we used $t \gg \beta$.

Because the answer we expect is proportional to ${1\over \rho(E)}$, it is convenient to go to the microcanonical spectral form factor, using
\begin{align}
\int {\d\beta\over 2\pi \i} e^{2\beta E}Z_{\text{Tr}}(\beta+it,b_1)Z_{\text{Tr}}(\beta-it,b_2) &\approx {1\over 4\pi t} \int {\d \beta\over 2\pi \i} e^{2\beta E}\exp\left\{\i\frac{b_1^2-b_2^2}{4t} - \frac{\beta}{4t^2}(b_1^2+b_2^2)\right\} \\
&={1\over 4\pi t} \exp\left\{\i\frac{b_1^2-b_2^2}{4t}\right\}\delta\left({b_1^2+b_2^2\over 4t^2}-2E \right).
\end{align}
We can also evaluate the integral over $a$, getting\footnote{We choose to include in $V$ the factor of two from the sum over orientation reversal of one of the trumpets.}
\be\label{eqn:v1/2_2}
V_{1/2}(b_1,b_2) = 2\times\int_\epsilon^{a_*}\frac{\d a}{2\tanh(\frac{a}{4})} = 2\log \cosh\frac{b_1+b_2}{4} + 2\log \cosh\frac{b_1-b_2}{4}  - 4\log\frac{\epsilon}{4}.
\ee
We can now evaluate the microcanonical spectral form factor for fixed $E$ and large $t$:
\begin{align}
K_E(t)&\supset \frac{e^{-S_0}}{4\pi t} \int_0^\infty b_1\d b_1 b_2\d b_2 V_{1/2}(b_1,b_2) \exp\left\{\i\frac{b_1^2-b_2^2}{4t}\right\}\delta\left({b_1^2+b_2^2\over 4t^2}-2E \right)\label{structure}\\
&\approx e^{-S_0} {2t^2\sqrt{E}\over \pi} \int_{-\infty}^\infty \d (\delta b) e^{\i \sqrt{E} \delta b} (\log \cosh(\sqrt{E}t)+\log \cosh{\delta b\over 4} - 2\log\frac{\epsilon}{4})\label{deltab}\\
&=e^{-S_0} {2t^2\sqrt{E}\over \pi} \int_{-\infty}^\infty \d (\delta b) e^{\i \sqrt{E} \delta b}\log \cosh{\delta b\over 4}\\
&=-{t^2\over 2\pi ^2 \rho(E)}.\label{expdecay}
\end{align}
which matches the encounter result (\ref{eqn:k1/2}). In the last step we used that in JT gravity, the density of states is $\rho_0(E) = \sinh(2\pi \sqrt{E})/(2\pi)^2$.

We will now make a few remarks connecting this calculation to the encounter picture (see appendix \ref{app:soft} for more details). The $b_1,b_2$ parameters can be regarded as analogous to the lengths of the periodic orbits, with difference of orbit actions $\Delta S = S_\gamma - S_{\gamma'}$ analogous to $(b_1^2-b_2^2)/4t \approx \sqrt{E}\delta b$. With this understanding we can write the moduli space volume in JT as
\be
V_{1/2}(b_1,b_2)  = 2\sqrt{E} t + 2\log\cosh\frac{\Delta S}{4\sqrt{E}} + \text{const.}
\ee
In periodic orbit theory, we should compare this moduli space volume to the integral over the parameters of the encounter with fixed action difference $\Delta S$:
\be
\int_{-c}^{c}\d s \d u \delta(su-\Delta S) {t-2t_{enc}\over t_{enc}}=\lambda\left(t-2t_{enc}\right)=\lambda t+2\log{|\Delta S| \over c^2}.
\ee
The Lyapunov exponent $\lambda$ from periodic orbit theory should be compared to the JT gravity chaos exponent $2\pi/\beta = 2\sqrt{E}$, so the terms linear in $t$ match. However, these terms drop out after integrating over $\Delta S$ with weighting $e^{\i \Delta S}$. Instead, the answer is determined by the subleading terms, and in particular by the locations of their singularities in the upper half-plane for $\Delta S$. In the JT case, the closest singularity to the real axis is at $\Delta S = 2\pi\i \sqrt{E}$, leading to the exponential suppression of (\ref{expdecay}).

\section{Beyond encounters}\label{SectionFour}
In a GUE-like theory such as orientable JT gravity, the analog of encounters are expected to cancel exactly at fixed energy. In this section, we will discuss a convenient decomposition of the moduli space that separates the contributions of different encounters. This will make it possible to understand the encounter contributions and their cancellation, as well as the failure of their cancellation at low energies.

Instead of JT gravity, we will work with the simpler Airy model, which may be viewed as the low energy or low temperature limit of JT gravity, where one approximates the $\sinh( c \sqrt{E})$ density of states as $\rho(E) = c\sqrt{E} $. In this limit, the lengths of the asymptotic boundaries, as well as the lengths of any internal closed geodesics, go to infinity. One can see this by taking this limit in the JT gravity formula for partition functions as trumpets integrated against the Weil-Petersson (WP) volume 
\be\label{JTpartitionfunctions}
\langle Z(\beta_1)\dots Z(\beta_n)\rangle_{\text{JT}}\supset e^{\chi S_0} \int_0^\infty b_1 \d b_1 \dots \int_0^\infty b_n \d b_n \; \frac{e^{-\frac{b_1^2}{4\beta_1}}}{\sqrt{4\pi \beta_1}} \dots \frac{e^{-\frac{b_n^2}{4\beta_n}}}{\sqrt{4\pi \beta_n}}  V_{g,n}(b_1,\dots, b_n).
\ee
Partition functions for the Airy model can be obtained from the JT answers by an infinite rescaling of $\beta$, accompanied by a renormalization of $S_0$
\be
\langle Z(\beta_1)\dots Z(\beta_n)\rangle_{\text{Airy}}= \lim_{\Lambda\rightarrow \infty} \Lambda^{\frac{3}{2}\chi} \langle Z(\Lambda \beta_1)\dots Z(\Lambda \beta_n)\rangle_{\text{JT}}
\ee
To take this limit in (\ref{JTpartitionfunctions}), we rescale the $b_i$ by $\sqrt{\Lambda}$. The WP volumes are polynomials in the $b_i$, with degree $6g+2n-6$. We define the Airy volumes as
\be
V_{g,n}^{\text{Airy}}(b_1\dots b_n) = \lim_{\Lambda\rightarrow \infty} \Lambda^{3-3g-n} V_{g,n}(\Lambda b_1,\dots,\Lambda b_n).
\ee
These Airy volumes are then homogeneous polynomials in the $b_i$ of degree $6g+2n-6$, given by the leading powers of the full WP volumes. The Airy partition functions can be written as trumpets integrated against the Airy volumes, with $S_0 \rightarrow S_0 +\frac{3}{2}\log(\Lambda)$.

In the limit where the boundary lengths $b_1\dots b_n$ become infinitely long, the surfaces counted by the WP volumes simplify. The Gauss-Bonnet theorem implies that a constant negative curvature surface with geodesic boundaries has a fixed volume proportional to its Euler character. As the lengths of the boundaries are going to infinity, the surfaces must become infinitely thin strips in order for the volume to remain fixed. 

This thin strip limit allows for a simple decomposition of the moduli space of these surfaces, described by Kontsevich \cite{kontsevich1992intersection}, which will connect in a transparent way to the encounters discussed in the previous section and to the description of the Airy model using a double-scaled matrix integral. We now briefly review this decomposition, following \cite{DoThesis}.

\subsection{Kontsevich's decomposition of moduli space}
In the thin strip (Airy) limit, the moduli space can be described as a sum over trivalent ribbon graphs, together with an integral over the  lengths of the edges that make up the graphs, subject to the constraint that the boundaries have lengths $b_i$:
\be
V^{\text{Airy}}_{g,n}(b_1\dots b_n) =\frac{2^{2g-2+n}}{|\text{Aut}(\Gamma)|}\prod_{k = 1}^E \int_0^{\infty} \d l_k \prod_{i=1}^n \delta(b_i - \sum_{k = 1}^E n^i_k l_{k})
\ee
Here $E = 6g-6+3n$ is the number of edges in the graph, $l_k$ is the length of edge $k$, and $n^i_k \in \{0,1,2\}$ is the number of sides of edge $k$ that belong to boundary $i$.

The Laplace transform of this expression is a little simpler:
\begin{align}\label{LapVMat}
\tilde{V}^{\text{Airy}}_{g,n}(z_1\dots z_n) &\equiv \int_0^\infty \prod_{i = 1}^n\left[\d b_i e^{-b_i z_i}\right] V^{\text{Airy}}_{g,n}(b_1\dots b_n) \\ &=\sum_{\Gamma \in \Gamma_{g,n}}\frac{2^{2g-2+n}}{|\text{Aut}(\Gamma)|} \prod_{k=1}^{6g-6+3n}\frac{1}{z_{l(k)}+z_{r(k)}}.
\end{align}
Here $\Gamma_{g,n}$ is set of trivalent ribbon graphs with genus $g$ and $n$ boundaries, contructed from $E=6g-6+3n$ edges and $V=4g-4+2n$ trivalent vertices. The $k$ variable runs over the $6g-6+3n$ edges, and the $l(k)\in \{1\cdots n \}$ index labels which boundary of the Riemann surface the left side of the ribbon belongs to. Similarly, $r(k)$ labels which boundary the right side of the ribbon belongs to.

We are interested in the case where there are two boundaries, so we will draw ribbon graphs with ribbon edges denoted by solid red ($1$) and dashed black ($2$) lines. Then a $11$ edge comes with a factor of $1/(2z_1)$, a $22$ edge comes with a factor of $1/(2z_2)$, and a $12$ edge comes with a factor of $1/(z_1+z_2)$. These ribbon graphs can be orientable or non-orientable, depending on what variety of JT gravity or Airy gravity we are interested in.\footnote{In the non-orientabe case, one also has additional factors of two due to the possibility of inserting orientation reversing operators along particular cycles.} An example in the non-orientable case  is
\be
\includegraphics[scale = .65, valign = c]{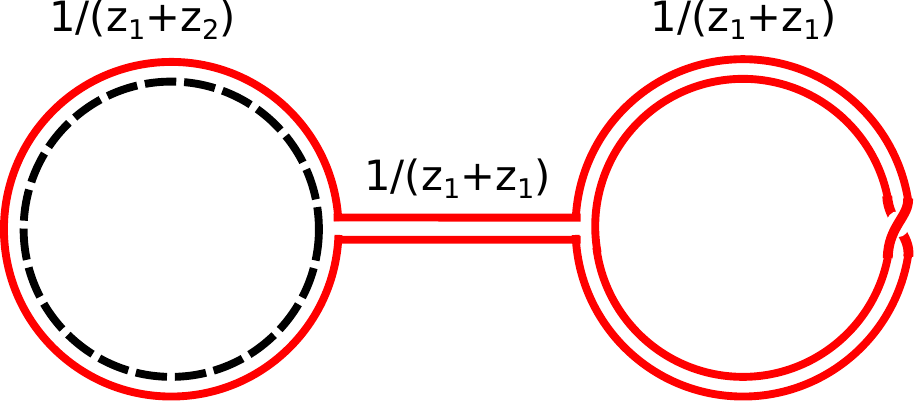}
\ee
This graph has two boundaries and genus one-half, and together with three other graphs discussed in section \ref{genusonehalfsubsection} below, it gives the Kontsevich-graph description of the Sieber-Richter two-encounter.

We will also consider the graphs with two boundaries and genus one. To enumerate the graphs, a useful fact is that all of the orientable graphs for fixed $(g,n)$ can be obtained from a single graph by repeatedly applying the cross operation (or Whitehead collapse) \cite{Whitehead1936equivalent,Penner1988perturbative}:
\be\label{crossOp}
\includegraphics[valign = c, scale = .7]{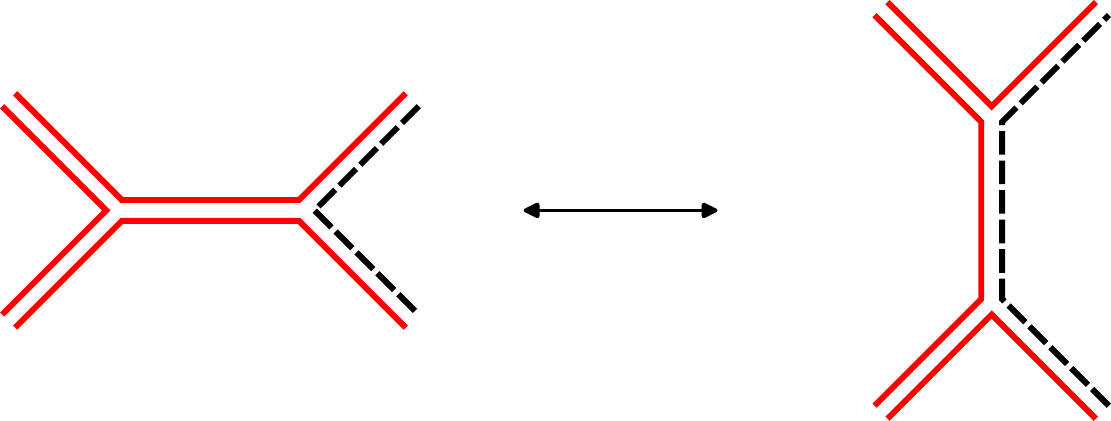}
\ee
This is consistent with that fact that moduli space $\overline{\mathcal{M}}_{g,n}$ is a connected space: if we back off of the Airy limit of JT gravity, then the strips have finite width, and the operation (\ref{crossOp}) is a smooth transition.

\subsection{Genus one-half}\label{genusonehalfsubsection}
In this section we will illustrate the connection between encounters and Kontsevich graphs by studying the example of genus one-half, with two boundaries. In this case, the volume of the moduli space is 
\be\label{Airygenusonehalfvol}
V^{\text{Airy}}_{\frac{1}{2},2}(b_1,b_2) = \text{Max}(b_1,b_2).
\ee
This can be obtained by taking the large $b_1,b_2$ limit of the JT gravity answer (\ref{eqn:v1/2_2}). To take this limit, one drops the constant piece and replaces $\log \cosh(\frac{b_1\pm b_2}{4})$ with $\frac{1}{4}|b_1\pm b_2|$.

There are four Kontsevich graphs with two boundaries and genus one half:
\be
\includegraphics[scale = .55,valign = c]{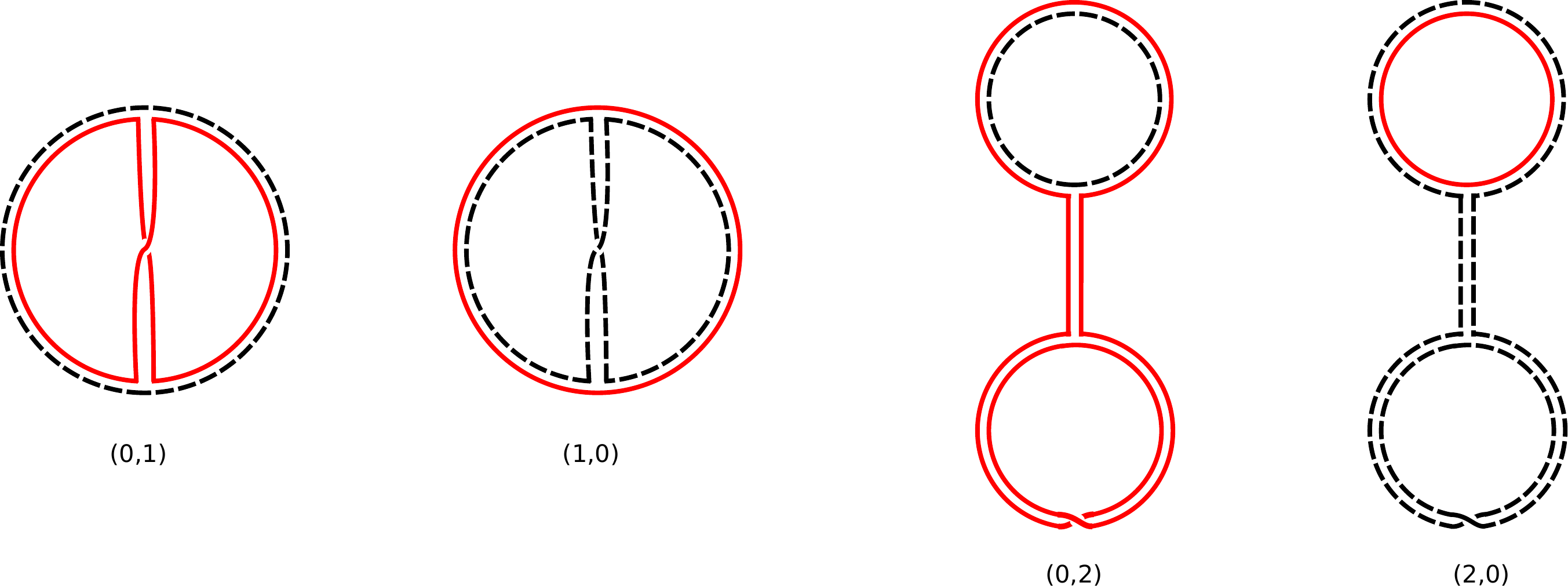}
\ee
Here the graphs are labeled by $(n_{11},n_{22})$, the number of $11$ and $22$ propagators. The contributions of these graphs to $\tilde{V}(z_1,z_2)$ are
\begin{align}
(1,0):\hspace{10pt} \frac{c_{1,0}}{z_1 (z_1+z_2)^2} &, \hspace{20pt} (0,1):\hspace{10pt} \frac{c_{0,1}}{z_2 (z_1+z_2)^2} \\
(2,0):\hspace{10pt} \frac{c_{2,0}}{z_1^2 (z_1+z_2)} &, \hspace{20pt} (0,2):\hspace{10pt} \frac{c_{0,2}}{z_2^2 (z_1+z_2)}
\end{align}
where the coefficients $c_{0,1} = c_{1,0}$ and $c_{0,2} = c_{2,0}$ are determined the by the symmetry factor of the graph, together with a factor of two from the possibility of orientation reversal along one boundary. 

Rather than computing the symmetry factors, we can compute $c_{1,0}$ and $c_{2,0}$ indirectly by matching to the volume (\ref{Airygenusonehalfvol}). To find the contribution of each graph to the volume, we take the inverse Laplace transform, for example
\be
(1,0):\hspace{10pt}  \int_{\gamma+i\mathbb{R}}\frac{\d z_1}{2\pi i} \frac{\d z_2}{2\pi i}e^{b_1 z_1 +b_2 z_2}  \frac{c_{1,0}}{z_1 (z_1+z_2)^2} = c_{1,0} b_2 \theta(b_1-b_2).
\ee
Together with a similar term from $(0,1)$, this gives
\be\label{genusonehalfvolsfirstgraph}
(1,0)+(0,1) = c_{1,0} \text{min}(b_1,b_2).
\ee
Similarly,
\be
(2,0) + (0,2) = c_{2,0} |b_1-b_2|.
\ee
To match to (\ref{Airygenusonehalfvol}) we conclude that $c_{1,0} = c_{2,0} = 1$.

The corresponding contributions to the spectral form factor
\be\label{b1b2integral}
\langle Z(\beta_1)Z(\beta_2)\rangle \supset e^{-S_0}\int \frac{b_1 \d b_1}{\sqrt{4\pi \beta_1}}\frac{b_2 \d b_2}{\sqrt{4\pi \beta_2}} e^{-\frac{b_1^2}{4\beta_1} - \frac{b_2^2}{4\beta_2}}V(b_1,b_2)
\ee
are then (keeping the leading power of $t$)
\begin{align}
(1,0) + (0,1) &= e^{-S_0}\frac{t^2}{\sqrt{2\pi\beta}}\\
(2,0) + (0,2) &= -2e^{-S_0}\frac{t^2}{\sqrt{2\pi\beta}}
\end{align}
The sum of these contributions is $-e^{-S_0}t^2/\sqrt{2\pi\beta}$, which is the Laplace transform of the microcanonical answer $-e^{-S_0} t^2/(\pi\sqrt{E})$, which matches the two-encounter contribution (\ref{eqn:k1/2}) in the special case of the Airy density of states $\rho(E)=\frac{\sqrt{E}}{2\pi}$. Of course, this follows from the low-energy limit of the match we previously found in JT gravity. The interesting feature is that both classes of graphs contribute at the same order, and we have to sum both in order to reproduce the answer from the encounter.

The $(1,0)$ and $(0,1)$ graphs naively resemble a two-encounter; if one shrinks away the $11$ (or $22$) propagator, we find a graph with only 12 propagators and a quartic vertex. The $12$ propagators correspond to nearly parallel stretches of the $1$ and $2$ geodesic boundaries on the surface, so these graphs represent contributions for which the two boundaries are nearly parallel in a pattern that matches the Sieber-Richter pair. Of course, in computing the spectral form factor one glues on trumpets to the surface with geodesic boundaries, but the asymptotic boundaries also remain almost parallel for the stretches. Along these stretches, the geometry locally looks like the double-cone (or a non-orientable ``twisted" double-cone).

The $(2,0)$ and $(0,2)$ graphs are not as obviously connected to encounter theory, but they do represent a small part of the moduli space integral that is analogous to the $s,u$ integration in the encounter. To see which part of moduli space it corresponds to, consider the $a$ geodesic. In the Airy limit, this is simply the shortest loop on the Kontsevich graph that includes the twisted edge. For the $(2,0)$ and $(0,2)$ graphs, this means that $a$ is the twisted edge itself, which forms a loop shorter than $|b_1-b_2|/2$. For the $(1,0)$ and $(0,1)$ graphs, $a$ corresponds to a loop that includes the twisted edge plus the shorter untwisted edge, with total length longer than $|b_1-b_2|/2$. So the two classes of graphs divide the moduli space up as 
\begin{align}
V^{(\text{Airy})}_{\frac{1}{2},2}(b_1,b_2) &= 2\int_0^{a^*=\frac{1}{2}\text{Max}(b_1,b_2)} \d a  \\ &= 2\underbrace{\int_{\frac{|\delta b|}{2}}^{a^*=\frac{1}{2}\text{Max}(b_1,b_2)} \d a}_{(1,0)+(0,1)= \frac{1}{2} \text{Min}(b_1,b_2)}+ 2\underbrace{\int_0^{\frac{|\delta b|}{2}}\d a}_{(2,0)+(0,2)=\frac{|\delta b|}{2}}.
\end{align}
By splitting the integral into two parts, we introduce ``fictitious" endpoint contributions, proportional to $|\delta b|$, which cancel between the two graphs.

We can understand the geometry a bit better by fattening the Kontsevich graphs up and connecting them to the embedding space diagram (\ref{embeddingdiagram}). Here we will focus on the part of the embedding diagram bounded by the $b_1$, $b_2$ geodesics, removing the asymptotic trumpets. The two classes of Kontsevich graphs correspond to two limiting embedding space diagrams, with the $a$, $a'$ geodesics shown:
\begin{equation*}\label{limitingdiagrams}
\includegraphics[valign = c, width = \textwidth]{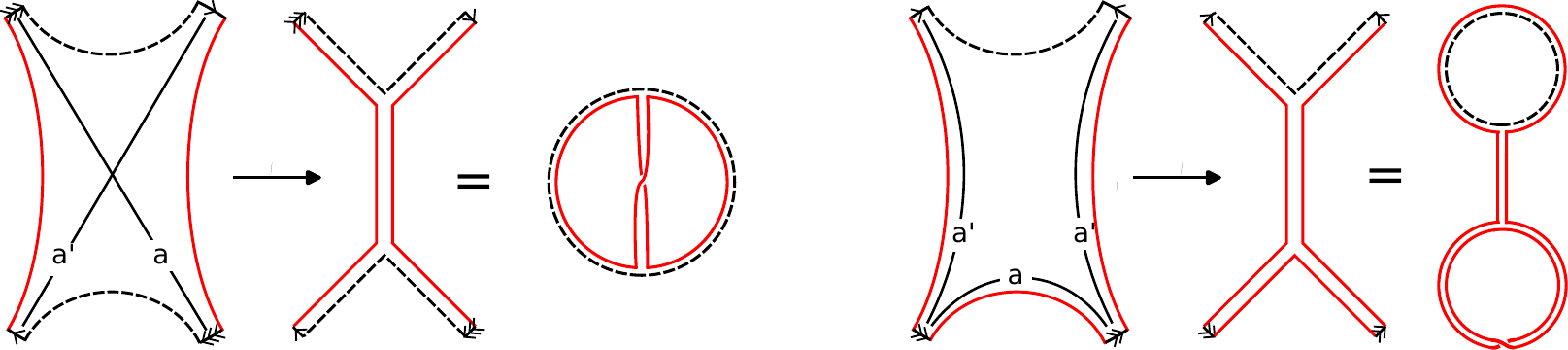}
\end{equation*}
On the left we start with a diagram similar to the middle of (\ref{embeddingdiagram}). A limiting case of this diagram represents a strip-like geometry. Upon making the identifications indicated by the arrows, we end up with the $(1,0)$ graph. On the right we begin with a somewhat different-looking embedding space diagram, which limits to the $(2,0)$ graph.

Though the two embedding diagrams that we start with look somewhat different, we can see that their topology is the same after making the indicated identifications. To see this more clearly, we may cut the embedding space diagram corresponding to the $(2,0)$ graph, then glue a pair of the identified edges to end up with an embedding space diagram resembling the $(1,0)$ diagram.
\be
\includegraphics[valign = c, scale = 1.2]{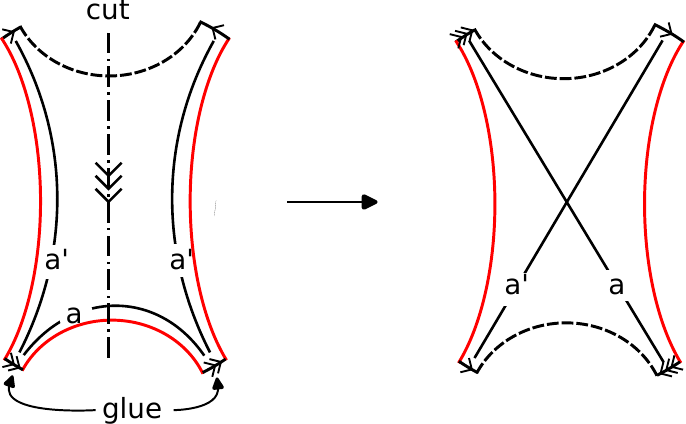}
\ee
After cutting and gluing the $(2,0)$, it must also be deformed somewhat to match the $(1,0)$ diagram; for instance, the newly cut geodesic, with an identification indicated by three arrows, is ``long" on the left diagram, but ``short" on the right. This corresponds to the fact that as shown, each of these two embedding space diagrams represent different limiting regions of moduli space, corresponding to the distinct $(1,0)$ and $(2,0)$ graphs. The limiting case of this deformation corresponds to the cross operation on the ``middle" edge of the $(2,0)$ graph.

\subsection{Genus one}
We now turn to our main interest, which is the first nontrivial ($\tau^3$) term in the series (\ref{airyanssec1}) for a GUE-like theory. This term arises at genus one. At genus one with two boundaries, the volume of the moduli space in the Airy limit is
\be\label{AIRY12}
V^{\text{Airy}}_{1,2}(b_1,b_2) = \frac{(b_1^2+b_2^2)^2}{192}.
\ee
Integrating this against trumpet wave functions and taking the limit of large $t$ leads to the term $-\tau^3/(6\pi)$ in the spectral form factor (\ref{airyanssec1}). We can gain a bit of insight by understanding how this contribution arises from different Kontsevich graphs, which can be related in turn to encounters.

\begin{figure}
\includegraphics[width = \textwidth]{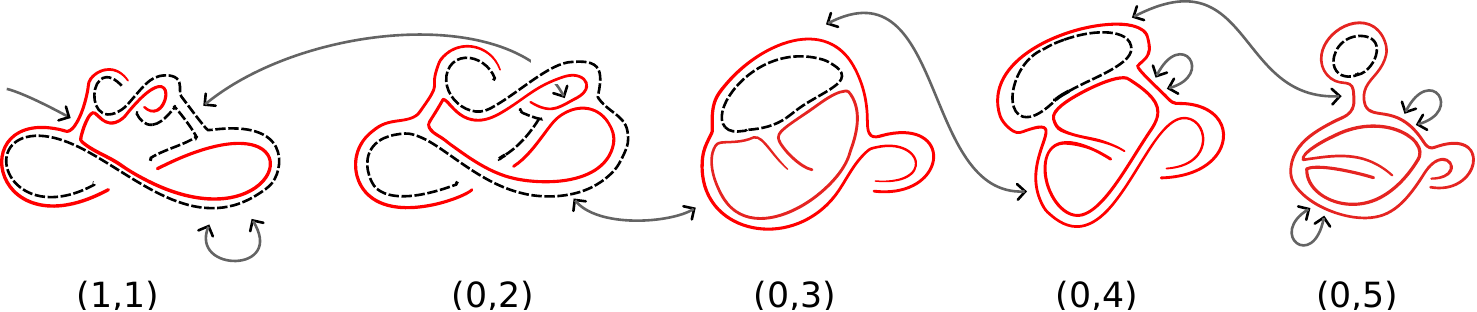}
\caption{{\sf \small The nine Kontsevich graphs with two boundaries and genus one consist of these five, together with another four given by interchanging the solid/red and dashed/black lines on the last four graphs. The dashed/black lines correspond to $1$ boundaries, and the solid/red lines correspond to $2$. The gray lines with arrows show what happens if we apply a cross operation to a given edge.}}\label{figGenusOneGraphs}
\end{figure}
The Kontsevich graphs that contribute to $V_{1,2}$ have six propagators total, which can be $11$, $22$ and $12$ propagators. Up to symmetries, there are nine distinct graphs, see Figure \ref{figGenusOneGraphs}, and they can be characterized by the number of $11$ and $22$ propagators, 
\be\label{k1k2}
(5,0), \ (4,0), \ (3,0), \  (2,0), \ (1,1), \ (0,2), \ (0,3), \ (0,4), \ (0,5).
\ee
For example, the $(5,0)$ graph has five $11$ propagators, zero $22$ propagators, and one $12$ propagator. It is given by 
\be
\frac{c_{5,0}}{z_1^5 (z_1+z_2)}
\ee
where the constant $c_{5,0}$ can be computed by evaluating the symmetry factor of the graph. As in genus one-half, these factors can be determined indirectly by matching to (\ref{AIRY12}). For example, after inverse Laplace transforming this, we find that the contribution to the volume $V^{\text{Airy}}_{1,2}(b_1,b_2)$ is 
\be
\int_{\gamma + \i \mathbb{R}}\frac{\d z_1}{2\pi \i}\frac{\d z_2}{2\pi\i} e^{b_1 z_1 + b_2 z_2} \frac{c_{0,5}}{z_1^5 (z_1+z_2)} =  \frac{c_{0,5}}{24}(b_1-b_2)^4\theta(b_1-b_2).
\ee
One can work out a similar expression for each of the $(k_1,k_2)$ cases in (\ref{k1k2}), and the coefficients $c_{k_1,k_2}$ are uniquely determined by the condition that the contributions of all of the graphs should add up to (\ref{AIRY12}). Explicitly, 
\be
c_{5,0} = c_{4,0} = \frac{1}{8}, \hspace{20pt} c_{3,0} = \frac{1}{6}, \hspace{20pt} c_{2,0} = \frac{1}{4}, \hspace{20pt} c_{1,1} = \frac{1}{2}
\ee
and equal values for $k_1\leftrightarrow k_2$.

The contribution of a given graph to the spectral form factor is then obtained from
\be\label{b1b2integral}
\langle Z(\beta_1)Z(\beta_2)\rangle \supset e^{-2S_0}\int \frac{b_1 \d b_1}{\sqrt{4\pi \beta_1}}\frac{b_2 \d b_2}{\sqrt{4\pi \beta_2}} e^{-\frac{b_1^2}{4\beta_1} - \frac{b_2^2}{4\beta_2}}\int_{\gamma + \i \mathbb{R}}\frac{\d z_1}{2\pi \i}\frac{\d z_2}{2\pi\i} e^{b_1 z_1 + b_2 z_2}\frac{c_{k_1,k_2}}{z_1^{k_1}z_2^{k_2}(z_1+z_2)^{6-k_1-k_2}}.
\ee
This integral reduces to a sum of hypergeometric functions. We can simplify the expression by setting $\beta_1 = \beta + \i t$ and $\beta_2 = \beta - \i t$ with large $t$, and keeping all terms that grow at order $t^3$ or faster. This leads to
\begin{align}
(5,0) + (0,5)  &= e^{-2S_0}\frac{t^3}{6\pi}\\
(4,0) + (0,4) &= e^{-2S_0}\frac{t^3}{6\pi}\left(3\log\frac{2t}{\beta} - 9\right)\\
(3,0) + (0,3) &= e^{-2S_0}\frac{t^3}{6\pi}\left(-6\log\frac{2t}{\beta}+10\right)\\
(2,0) + (0,2) &= e^{-2S_0}\frac{t^3}{6\pi}\left(-\frac{t^2}{\beta^2} + 3\log\frac{2t}{\beta}-3\right)\label{02}\\
(1,1) &= e^{-2S_0}\frac{t^3}{6\pi}\cdot\frac{t^2}{\beta^2}\label{11}
\end{align}
The sum of these contributions gives $-e^{-2S_0}t^3/(6\pi) = - e^{S_0}\tau^3/(6\pi)$, which produces the cubic term in (\ref{airyanssec1}). However, individual graphs contain terms that grow faster with time.

\subsubsection{Encounters}
As in the genus one-half case from section \ref{genusonehalfsubsection}, we can make a map from Kontsevich graphs to encounters by shrinking the $11$ and $22$ edges to form a graph with only $12$ edges but with higher-degree vertices. If we do this, the $(1,1)$ and $(0,2)$ graphs will correspond to a case with two two-encounters, and the $(0,3)$ and $(0,4)$ graphs will correspond to a three-encounter. After shrinking the $22$ edges, the final graph $(0,5)$ does not correspond to an encounter, but in parallel to the discussion of the $(0,2)$ graph from genus one-half, we believe it should be considered part of the extended three-encounter moduli space.

Let's examine the Kontsevich graphs that correspond to a pair of two-encounters. The contribution is the sum of (\ref{02}) and (\ref{11}), which gives
\be
(2,0) + (0,2) + (1,1) = e^{-2S_0}\frac{t^3}{2\pi}\left(\log\frac{2t}{\beta} - 1\right).\label{graph2enc}
\ee
We would like to compare this to the semiclassical answer for a pair of two-encounters (\ref{twoEncounters}), for the density of states of the Airy model $\rho(E) = \sqrt{E}/(2\pi)$
\begin{align}
\text{two two-encounters} &=  e^{-2S_0}\frac{t^3}{2\pi E}.
\end{align}
In the canonical ensemble, this gives the naive expression
\be
\text{two two-encounters} \stackrel{?}{=} e^{-2S_0}\frac{t^3}{2\pi}\int_0^\infty\frac{\d E}{E} e^{-2\beta E}.
\ee
The reason this expression is naive is that at very low energies, the encounter picture breaks down, because the action is small enough that we do not require orbits to form pairs whose action cancels. 

For the case of genus one-half, this breakdown was not significant because the analogous integral over energy was $\int \d E  e^{-2\beta E}/ \sqrt{E}$ which is convergent. But in the present case, the integral diverges and the cutoff associated to the breakdown of encounter theory becomes important. We can estimate the energy of the breakdown from the point where the action $S \sim E t$ becomes of order one, which gives $E \sim 1/t$. A revised estimate for the semiclassical encounter contribution would then be
\be\label{twotwoencounters}
\text{two two-encounters} \stackrel{?}{=} e^{-2S_0}\frac{t^3}{2\pi}\int_{1/t}^\infty\frac{\d E}{E} e^{-2\beta E} = e^{-2S_0}\frac{t^3}{2\pi}\left(\log\frac{t}{\beta} +\text{const}\right).
\ee
This matches the form of (\ref{graph2enc}).

One can similarly find agreement between the predicted contribution of a three enounter and the sum of the graphs $(5,0) + (0,5) + (4,0) + (0,4) + (3,0) + (0,3)$. In particular, the cancellation of the encounters demonstrated in \cite{muller2005periodic} is visible here in the fact that the log terms cancel between the graphs summed in (\ref{graph2enc}) and the three-encounter graphs. 

\subsubsection{Beyond encounters}
Because the $t^3\log(t)$ terms cancel, the entire contribution comes from the $t^3$ terms, and in encounter language, these contributions depend on the details of the small energy region (e.g.~the precise cutoff one uses in (\ref{twotwoencounters})). This cannot be computed using standard encounter theory. However, the Kontsevich graphs continue to be valid for all energies. In this sense, the Kontsevich graphs give a quantum completion of the semiclassical encounter theory for this particular system.

It is interesting to understand the region of the $b_1,b_2$ integral (\ref{b1b2integral}) that is relevant for the $t^3\log(t)$ pieces that cancel out vs.~the full $t^3$ answer. The log terms arise from non-analyticities at $b_1 = b_2$, where the phases contributed by the trumpet wave functions cancel. This is analogous to the fact that encounter contributions in periodic-orbit theory arise from a nonanlyticity in the region $\Delta S = 0$ where a pair of orbits have approximately the same length and cancelling actions.

However, the full moduli space volume is analytic in $b_1,b_2$, which implies that the log terms must cancel when we sum over graphs. What region of the $b_1,b_2$ integral is important for producing the leftover $t^3$? We have an integral of the form
\be
\frac{1}{t}\int_0^\infty b_1 \d b_1 b_2 \d b_2 e^{\i (b_1^2-b_2^2)/(4t)} (b_1^2-b_2^2)^2.
\ee
In this integral, $b_1^2\sim t$ and $b_2^2\sim t$, but with no particular preference for the region where $b_1 = b_2$. So the $1$ and $2$ boundaries have significantly different lengths, and the $11$ or $22$ portions of the Kontsevich graphs are as long as the $12$ portions. This corresponds to the idea that we are probing low energies, so that action $b_1^2/t$ is of order one and it does not need to cancel between the two ``orbits.'' Note that in periodic orbits, the analog of this region would be outside the regime of the validity of the semiclassical encounter approximations.

\section{Discussion}\label{Discussion}

In the Airy model at genus one, we found that the answer for $K_\beta(t)$ came from an integral over a large portion of moduli space. This poses a challenge for understanding the geometric origin of the series for $K_\beta(t)$ at higher genus and for theories with other densties of states; $K_\beta(t)$ has a universal form, fixed entirely by $\rho_0(E)$, but this universal answer comes from a highly quantum integral over moduli space. This suggests that there is some universal structure in the moduli space responsible for this series. Though we have not understood this structure, our findings in the Airy model hint at a relationship with encounters. 

 In the Airy model the moduli space has a natural structure, given by the Konstsevich graphs. In a sense we made precise at genus one half and genus one, the whole moduli space should be thought of as made up of ``quantum corrected" encounters, valid at very low energies. Perhaps in JT (and even in more general large $N$ chaotic systems) there is a ``fattened" version of this quantum encounter region of moduli space that is responsible for the answer, rather than the entire moduli space.

We can see a hint that the connection between the genus expansion for $K_\beta(t)$ and encounters generalizes to higher genus/other spectral curves by generalizing the estimate (\ref{twotwoencounters}) of $K_\beta(t)$ from encounters. An encounter configuration is expected to give a contribution to $K_E(t)$ proportional to $t^{2g+1}/\rho(E)^{2g}$. An estimated contribution of the encounter to $K_\beta(t)$, generalizing (\ref{twotwoencounters}) and extrapolating to low energies, would then be\footnote{We have dropped all terms that would be small in the $\tau$-scaling limit.}
\begin{align}\label{highergenuscutoff}
K_\beta(t) & \stackrel{?}{\supset}  C e^{-2g S_0} t^{2g+1} \int_{\frac{1}{t}}^{\infty} \frac{\d E}{\rho_0(E)^{2g}} e^{-2\beta E},
\cr
&=  C e^{-2g S_0} t^{2g+1} \bigg[P_g^{(\rho)}(\beta) \bigg(\log\frac{t}{\beta}+\text{const}\bigg) + \text{Higher powers of t}\bigg].
\end{align}
Here $P_{g}^{(\rho)}(\beta)$ is a polynomial in $\beta$ of degree $g-1$, whose coefficients depend on the first $g$ coefficients in the expansion (\ref{generaldensity}) for $\rho_0(E)$. 

Summing over encounters at each genus, the familiar cancellations between encounters in $K_E(t)$ imply that the log terms cancel, leaving us with cutoff-dependent terms which may or may not cancel between encounters. We can compare this estimate with the conjecture (\ref{tauexpansionintro}) for $K_\beta(t)$
\be\label{Kbetaexpansion}
K_\beta(t) = \sum_{g=0}^\infty P^{(\rho)}_{g}(\beta) \; e^{-2 g S_0}\; t^{2g+1}.
\ee
In appendix \ref{Pappendix} we show that the polyomials $P^{(\rho)}_{g}(\beta)$ in (\ref{highergenuscutoff}) and (\ref{Kbetaexpansion}) are indeed the same.\footnote{Up to an overall genus-dependent coefficient which can be absorbed into the coefficient $C$ in (\ref{highergenuscutoff}).} So the ``const" terms in the estimate (\ref{highergenuscutoff}) match the genus g contribution in (\ref{Kbetaexpansion}), up to an overall cutoff-dependent factor. 

Another set of questions concerns the relationship between the genus expansion for $K_\beta(t)$ and other approaches to understanding the plateau, such as the sigma model approach \cite{Wegner:1979tu,doi:10.1080/00018738300101531,PhysRevLett.75.902}, the Riemann-Siegel lookalike formula \cite{Berry_1990,Keating:1992tq,BerryKeatin1992}, and orbit action correlation functions \cite{argaman1993correlations}. Understanding the relationship between these approaches and the approach taken in this paper may be useful for learning lessons about theories that do not have a $\tau$-scaled spectral form factor.  We discuss the sigma model approach in \hyperref[Appendix A]{Appendix A} and the action correlation approach in \hyperref[Appendix B]{Appendix B}.

\section*{Acknowledgements} 

We thank Alexander Altland, Adel Rahman, Julian Sonner and the authors of \cite{Blommaert:2022lbh,Weber:2022sov} for discussions and Raghu Mahajan and Stephen Shenker for initial collaboration. PS is supported by a grant from the Simons Foundation (385600, PS), and by NSF grant PHY-2207584. DS is supported in part by DOE grant DE-SC0021085 and by the Sloan Foundation. ZY is supported in part by the Simons Foundation.

\appendix

\section{Airy sigma model}\label{Appendix A}
In this appendix we will use the  ``sigma model" approach to quantum chaos \cite{Wegner:1979tu,doi:10.1080/00018738300101531} to study the genus expansion for the plateau, following \cite{Haake:1315494}.\footnote{For a review which makes contact with two-dimensional gravity see \cite{Altland:2020ccq}.} The sigma model is closely related to encounters, and this approach will give us some perspective on how the encuonters are  ``regulated" in $K_\beta(t)$.

The basic object one considers in the sigma model approach is the generating function
\be\label{generatingfunction}
Z(E_1,E_2,E_3^+,E_4^-) \equiv \bigg\langle\frac{\det(E_1-H)\det(E_2-H)}{\det(E_3+i \epsilon -H)\det(E_4-i\epsilon-H)}\bigg\rangle_{H}.
\ee
Here $\langle \cdot \rangle_H$ denote averaging over an ensemble of Hamiltonians $H$. From the generating function, one can extract the pair correlator of resolvents,
\be\label{resolventfromgenerating}
\big\langle R_+(E_1) R_-(E_2)\big\rangle_H = \partial_{E_3}\partial_{E_4} Z(E_1,E_2,E_3^+,E_4^-)\big|_{E_3\rightarrow E_1,E_4\rightarrow E_2},
\ee
and from this, the pair density correlator $\langle \rho(E_1)\rho(E_2)\rangle_{H} = \frac{1}{\pi^2}\text{Re}\langle R_+(E_1) R_-(E_2)\rangle_H$.\footnote{For an ordinary matrix integral the tree-level resolvent has a real part which does not contribute to the density correlator. However in the Airy model and other double-scaled matrix integrals, we redefine the resolvent with this part subtracted off. See \cite{Saad:2019lba} for more detail.} To obtain the density correlator, it was important that we gave $E_3$ and $E_4$ infinitesimal imaginary parts of opposite sign. This difference in sign, or  ``causality", plays a key role in this approach. 

In the Airy model, we can represent $Z(E_1,E_2,E_3^+,E_4^-)$ as an integral over a $4\times 4$ Hermitian supermatrix $A_{ab}$, with indices $a=1\dots 4$ corresponding to one of the four determinants in (\ref{generatingfunction}). We assign each determinant a grading (fermionic for the determinants in the numerator, bosonic for those in the denominator), and a  ``causality", related to the sign of the infinitesimal imaginary part of the energy. The determinants with energies $E_1,E_3$ are assigned an advanced causality, while the determinants with energies $E_2,E_4$ are assigned a retarded causality. For a more detailed derivation of the analogous integral for $\langle \frac{\det(E_1-H)}{\det(E_2-H)}\rangle$ in the Airy model, see appendix A.1 of \cite{Saad:2019lba}.

Explictly, the matrix integral is
\be\label{SUSYkontsevich}
Z(E_1,E_2,E_3^+,E_4^-) = \int \d A \exp\bigg\{ e^{S_0}\bigg(\frac{4}{3}\STr[A^3]+ \STr[\hat{E} A]\bigg)\bigg\}.
\ee
Here $\hat{E} = \text{diag}(E_1\dots E_4)$. $\STr[\cdot]$ denotes the supertrace of a supermatrix; for a supermatrix $S$ with a bosonic-bosonic block $S_b$ and a fermionic-fermionic block $S_f$, $\STr[S]=\Tr[S_b]-\Tr[S_f]$.

The integral  (\ref{SUSYkontsevich}) is a supersymmetric generalization of the Kontsevich integral, and the double-line diagrams of this cubic matrix integral are closely related to the trivalent ribbon graphs used in Section \ref{SectionFour}.

For $E_1=E_2=E_3=E_4=E$, the integral has a $U(2|2)$ symmetry $A\rightarrow T A T^{-1}$, $T\in U(2|2)$. To compute the resolvent, we take derivatives of  (\ref{SUSYkontsevich}) and set $E_3=E_1$, $E_4=E_2$. Then for $\delta E= E_1-E_2$ small, there is an exact $U(1|1)\times U(1|1)$ subgroup which remains an exact symmetry, with the remaining $\frac{U(2|2)}{U(1|1)\times U(1|1)}$  ``causal symmetry" explicitly but weakly broken.

As $\delta E\rightarrow 0$ the causal symmetry is spontaneously broken by the infinitesimal imaginary parts of the energies. Here this can be seen by doing a saddle point expansion around $\delta E = 2i \epsilon$. The infinitesimal imaginary energy difference picks out a particular solution $A_s= i \frac{\sqrt{E}}{2}\Lambda$, where $\Lambda$ is a diagonal matrix with entries $+1$ for indices of advanced causality, and $-1$ for indices of retarded causality. Then for small but finite $\delta E$, we can focus on a (pseudo-)Goldstone modes $A=i \frac{\sqrt{E}}{2}T\Lambda T^{-1}\equiv i \frac{\sqrt{E}}{2} Q$, $T\in U(2|2)$. $Q$ parametrizes the goldstone manifold $\frac{U(2|2)}{U(1|1)\times U(1|1)}$.

Restricting  (\ref{SUSYkontsevich}) to the pseudo-Goldstone manifold, and using (\ref{resolventfromgenerating}), we arrive at the effective  ``sigma model" computing the double-resolvent
\be\label{airysigmamodel}
-e^{2S_0}\frac{E}{4}\int \d Q \; e^{-I_{eff}[Q]}\;\STr[Q(P^+\otimes P^f)]\STr[Q(P^-\otimes P^f)],
\ee
\be\label{airysigmamodelaction}
I_{eff}[Q] = -i \delta E \frac{e^{S_0}\sqrt{E}}{4} \STr[Q\Lambda].
\ee
Here the measure for $Q$ is the Haar measure on $\frac{U(2|2)}{U(1|1)\times U(1|1)}$. 

It is important at this stage to note that there are other modes which become soft as $E_1,E_2\rightarrow 0$, rather than as $\delta E \rightarrow 0$. These modes are responsible for the singularities in the resolvent as $E_1, E_2\rightarrow 0$, and are related to the $11$ and $22$ propagators in the Kontsevich graphs. For $E\neq 0$ these modes have action proportional to $e^{S_0}$. On the other hand, the pseudo-Goldstone modes we have focused on have an action proportional to $\delta E e^{S_0} \equiv s$. $s$ is conjugate to $\tau$, rather than $t$, so in a fixed-energy version of the $\tau$-scaled limit we should hold $s$ fixed as $e^{S_0}\rightarrow\infty$. Then in the $\tau$-scaled limit, only the pseudo-Goldstone mode survives, and the effective sigma model becomes exact. In this limit it is useful to use a rescaled version of the double-resolvent $R_E(s)\equiv e^{-2S_0} \langle R_+(E+e^{-S_0}\frac{s}{2})R_-(E-e^{-S_0}\frac{s}{2})\rangle_{c,H}$, where we focus on the connected part of the double-resolvent.

Let's now review some results from the sigma model, in the specific case of the Airy model.

\begin{itemize}
\item One can explictly do the integral (\ref{airysigmamodel}) to find the result 
\be\label{Airydoubleresolvent}
R_E(s) = \frac{-1 + e^{ i s \sqrt{E}}}{2 s^2}.
\ee
where $s$ has a small positive imaginary part.

This simple form is a consequence of the fact that the integral (\ref{airysigmamodel}) is one-loop exact, with two saddle points. The goldstone manifold $\frac{U(2|2)}{U(1|1)\times U(1|1)}$ has a bosonic component $H^2\times S^2$. After integrating out the grassman variables, the measure for the bosonic variables is the natural measure on $H^2\times S^2$. The action (\ref{airysigmamodelaction}) decomposes into a sum of an $H^2$ term and an $S^2$ term. The $H^2$ integral has a single saddle point. The action for the $S^2$ component is proportional to the height function $\cos(\theta)$ on the sphere, and so there are two saddle points on opposite poles of the sphere.

\item The leading saddle\footnote{When $\delta E$, or  ``ramp" saddle, has a small imaginary part.} is given by $Q=\Lambda$, and contributes $\frac{-1}{2 s^2}$. The saddle point has zero action, and the quadratic fluctuations around the saddle give the ramp contribution to the double resolvent. 

Though the perturbative series trucates at one-loop it is interesting to see another perspective on how the higher-order corrections cancel. It turns out that the perturbative series can be precicely mapped onto the encounter expansion. To do this, one introduces the  ``rational parametrization" of $Q$,
\be
Q= T \Lambda T^{-1}, \hspace{20pt} T=\begin{pmatrix} 1 & B \cr \tilde{B} & 1 \end{pmatrix},
\ee
where $B,\tilde{B}$ are $2\times 2$ supermatrices in the advanced-retarded and retarded-advanced sectors respectively. In terms of these variables the action (\ref{airysigmamodelaction}) is
\be\label{sigmaencounterexpansion}
I_{eff} = -  i   s\frac{\sqrt{E}}{2}  \sum_{l=1}^\infty \STr[(B\tilde{B})^l].
\ee
The degree $l$ vertex for $l\geq 2$ corresponds to an $l$-encounter, with the Gaussian integral from the $l=1$ term giving Wick contractions that connect these encounters in all possible ways. This correspondence is precise; the contribution to the resolvent from a given set of these vertices precisely matches the contribution from the sum over encounter structures with the same number of encounters of a given degree.

\item The other saddle point, known as the Andreev-Altshuler (AA) saddle \cite{PhysRevLett.75.902} contributes $\frac{e^{i s \sqrt{E}}}{2s^2}$.\footnote{Explicitly, $Q_{AA}=\text{diag}(1,-1)\otimes \text{diag}(1,-1)$, where the first factor corresponds to the advanced/retarded sector and the second corresponds to the boson/fermion sector.} This saddle gives the rapid oscillations and coincident energy delta function in the density pair correlator, which are responsible for the plateau.

\item The action is proportional to the product $s\sqrt{E}$. So the pseudo-Goldstone mode becomes soft, and the nonperturbative saddle point becomes important, as either $s\rightarrow 0$ or $E\rightarrow 0$. This reflects the fact that $K_E(\tau)$ reaches the plateau at early times for small energies $E$.
\end{itemize}

To make contact with the genus expansion for the plateau, we introduce a version of the resolvent in the canonical ensemble,
\be
R_\beta(s)\equiv \int_0^\infty \d E \;e^{-2\beta E} R_E(s).
\ee
This is related to the spectral form factor through $K_\beta(\tau)=\frac{e^{S_0}}{\pi^2}\int ds \; e^{i s \tau} \text{Re}[R_\beta(s)]$. Using our expression (\ref{Airydoubleresolvent}) for $R_E(s)$ we can explicitly compute $R_\beta(s)$. The energy integral of the contribution of the ramp saddle point is trivial; the AA saddle point gives a contribution with a nontrivial asymptotic series around $s=\infty$, which we can identify with the genus expansion:
\be\label{rbetaseries}
R_\beta(s) \sim -\frac{1}{4\beta s^2}- \sum_{g=1}^\infty \frac{2^{2g-2} (2g-1)!! \beta^{(g-1)}}{ s^{2g+2}}.
\ee
One can verify that this genus expansion matches (\ref{airyanssec1}) upon Fourier transforming and dividing by $\pi^2$.

We can also write a version of the sigma model which computes $R_\beta(s)$ directly, by multiplying (\ref{airysigmamodel}) by $e^{-2\beta E}$ and integrating over $E$ before doing the $Q$ integral. We are left with an integral $\int dQ e^{-I^{\beta}_{eff}[Q]}\;\STr[Q(P^+\otimes P^f)]\STr[Q(P^-\otimes P^f)]$, where the new weighting factor $e^{-I^{(\beta)}_{eff}[Q]}$ is given by
\be
e^{-I^{\beta}_{eff}[Q]}= \int_0^\infty \d E\;e^{-2\beta E} e^{-I_{eff}[Q]}.
\ee
It is useful to understand how to derive the asymptotic series (\ref{rbetaseries}) from this version of the sigma model. Rather than doing this directly, we will formulate a toy version of this problem by starting with the sigma model computing $R_E(s)$, integrating out the grassman and $H^2$ degrees of freedom in $Q$, leaving the integral over the $S^2$ degrees of freedom, and then compute $R_\beta(s)$ by doing the energy integral before the $S^2$ integral. We find the integral
\be
R_\beta(s) = \frac{1}{2\pi s \beta^{3/2}}\int_{S^2} \d^2 x f\bigg(\frac{(1-I_{S^2}(x))s}{\sqrt{\beta}}\bigg).
\ee
$I_{S^2}(x)$ is the contribution to $\STr[Q \Lambda]$ from the $S^2$ variables. The path integral weight $f$ is a function of the combination $\frac{(1-I_{S^2})s}{\sqrt{\beta}}$.\footnote{Explicitly, $f(y)=-\frac{y }{64}-\frac{\sqrt{2\pi}}{512}(y^2-16) e^{-\frac{y^2}{32}}\big(\text{Erfi}(-\frac{y}{\sqrt{32}})+i\big)$}

The most straightforward way to evaluate the integral is to parametrize the $S^2$ using an azimuthal angle $\phi$ and a height function $h=\cos(\theta)$, where $\theta$ is the polar angle. Both are integrated with a flat measure. $I_{S^2}=h$, so we can trivially integrate out $\phi$.

The remaining integral over $h$ is
\be
\frac{1}{\beta^{\frac{3}{2}}s } \int_{-1}^{1}\d h \;f\bigg(\frac{(1-h)s}{\sqrt{\beta}}\bigg).
\ee
$f\big(\frac{(1-h)s}{\sqrt{\beta}}\big)$ has an asymptotic expansion around $s=\infty$, so one might attempt to reproduce the series (\ref{rbetaseries}) by performing the $h$ integral term by term. However, this expansion is not valid for $h$ near $1$ (corresponding to the location of the ramp saddle point), and the resulting integrals do not converge. Instead, we rewrite this integral as the difference
\be
\frac{1}{\beta^{\frac{3}{2}}s }  \int_{-\infty}^{1}\d h \;f\bigg(\frac{(1-h)s}{\sqrt{\beta}}\bigg) - \frac{1}{\beta^{\frac{3}{2}}s } \int_{-\infty}^{-1}\d h \;f\bigg(\frac{(1-h)s}{\sqrt{\beta}}\bigg).
\ee
Changing variables in the first term to $\tilde{h}=\frac{(1-h)s}{\sqrt{\beta}}$, we find the ramp contribution
\be
\frac{1}{\beta s^2} \int_{0}^\infty \d \tilde{h} \; f(\tilde{h}) = -\frac{1}{4\beta s^2}.
\ee 
In the second term the argument of $f$ is always large for large $s$, so we can expand around $s=\infty$ and do the $h$ integral term by term, yielding the second term in (\ref{rbetaseries}).

Here we could see clearly that the asymptotic series did not come from the region of the integral near the ramp saddle point. If we instead attempted to expand around the $h=1$ ramp saddle point and mimic the encounter perturbation theory, we find that the individual terms in this perturbation theory do not converge. 

In our toy version of the fixed-$\beta$ sigma model, restricted to the $S^2$ variables, the encounter expansion corresponds to changing variables from $h,\phi$ to a complex variable $z$ with $h=\frac{1-|z|^2}{1+|z|^2}$, $z= |z|e^{i\phi}$. $z$ is related to the fermion-fermion components of the matrices $B,\tilde{B}$ used in the conventional encounter expansion of the sigma model.\footnote{More precisely, $-|z|^2$ is an eigenvalue of $B\tilde{B}$ and $\tilde{B}B$.} The ramp saddle point is at $z=0$, and the AA saddle point is at $z=\infty$.

To attempt an analog of the encounter expansion, we first write $h=1+ 2\sum_{l=1}^\infty (-|z|^2)^l$. The degree $2 l$ vertices for $l\geq 2$ are analogous to the contribution of an $l$-encounter. To keep track of these, we introduce arbitary coefficients $h_l$, $l\geq 2$, which in the end we set equal to one, and replace $h\rightarrow h(z)= 1- 2 |z|^2 +2\sum_{l=2}^\infty h_l (-|z|^2)^l$ to find\footnote{Here we have modified the measure as well as the action. If one modifies the expansion (\ref{sigmaencounterexpansion}) to include coefficients $g_l$, the measure for the remaining bosonic modes resulting from the Grassman integrals would involve the $g_l$. The measure in (\ref{modifiedresolvent}) is this modified measure restricted with the $H^2$ variables set to zero.}
\be\label{modifiedresolvent}
R_\beta(s)= \frac{1}{2\pi \beta^{\frac{3}{2}}s} \int \d^2 z \; \big(1+\sum_{l=2}^\infty h_l (-|z|^2)^{l-1}\big)^2 f\bigg(\frac{2s}{\sqrt{\beta}}\big( |z|^2 +\sum_{l=2}^\infty h_l (-|z|^2)^l\big)\bigg)\bigg|_{h_l\rightarrow 1}.
\ee
However, if we expand the integrand in a power series in the $h_l$, the integral over $z$ of every term with at least one power of $g_l$ does not converge. $f\big(\frac{2s}{\sqrt{\beta}}|z|^2\big)$ decays slowly at infinity, as $\frac{1}{|z|^6}$. Each  ``encounter" comes with a power of at least $|z|^4$, so these terms decay at most as $\frac{1}{|z|^2}$ at infinity (with most terms in the expansion growing at infinity). The divergence of the encounter contributions corresponds to the fact that the encounter contributions diverge near $E=0$, but in this computation we are integrating the encounters over energy all the way down to $E=0$. 

A better method of doing the integral would be to change variables from $z,z^*$ to $h(z)$ and an angular variable. Each term in the large $s$ asymptotic series for fixed $h_l$ (such that the integral converges) would be sensitive to the large $z$ behavior of $h(z)$, and thus sensitive to all of the $h_l$. So in a sense, to compute a given term in the series we need to  ``resum" the encounters before doing the final integral over $|z|$ or $h(z)$.

The relationship between the sigma model computation of $R_\beta(s)$ and the gravity computations in section \ref{SectionFour} using the Kontsevich graphs is somewhat puzzling. In the gravity computations, the encounters were regulated at low energies, removing the divergences from $E=0$. The entire answer came from low-energy contributions, with $E\sim \frac{1}{t}$. 

In this sigma model computation, we took the limit $t\propto e^{S_0}\rightarrow \infty$ before doing the integral. For finite $E$ this freezes the modes in the integral  (\ref{SUSYkontsevich}) transverse to the pseudo-Goldstone manifold. These transverse modes are responsible for the low-energy corrections to encounters in the Kontsevich graph approach, but in this limit the energy at which they become important is sent to zero. In the sigma model we simply restrict to the pseudo-Goldstone modes, without keeping any zero-energy contributions from these modes.

These two approaches appear to be inconsistent with each other. In the sigma model approach, we could point to a possible ``mistake" in the computation: throwing away any zero-energy contributions. However, in the sigma model approach we still get the correct answer, with the contribution from low energies being small. This is an example of the main puzzle discused in the \hyperref[Discussion]{Discussion}.

\section{Action correlation in the Airy model}\label{Appendix B}
In periodic orbit theory, there is another perspective of the ramp-plateau transition based on the idea of an ``action correlation function" \cite{argaman1993correlations}.
In this appendix, we will study the analog of the ``action correlation function" in Airy model.
In the semiclassical limit $(E\gg 1)$, the Airy SFF can be written as a fourier transformation of the summation of Airy volumes:
\bea
	K_E(t)&=&{1\over 4\pi t} \int_0^{\infty} b_1 \d b_1 b_2 \d b_2 \exp\left(\i{b_1^2-b_2^2\over 4 t}\right)\delta\left({b_1^2+b_2^2\over 4t^2}-2E\right) \sum_g e^{-2g S_0}V^{\text{Airy}}_g(b_1,b_2)~~~\\
	&=&{1\over 4\pi \sqrt{E}}\int dx e^{2\i \sqrt{E}x} f(2t\sqrt{E}+x,2t\sqrt{E}-x).
\eea
Here $b_{1,2}$ are the lengths of the trumpet geodesics, and we define the orbit pair correlation function $f(b_1,b_2)$ as a sum over all the Airy volumes with two geodesic boundaries:
\be 
f(b_1,b_2)=b_1b_2\sum_g e^{-2g S_0}V^{\text{Airy}}_g(b_1,b_2).
\ee 
As discussed in section \ref{section:SRJT}, $b_{1,2}$ can be thought of as the analog of the lengths of periodic orbits in the microscopic theory.
Then the function $f(b_1,b_2)$ describes the joint probability of having two periodic orbits with lengths $b_1,b_2$.
In the $\tau$-scaled limit, the ramp-plateau transition is related to the pair correlation function with large average length $b_1\sim b_2\sim t\sim e^{S_0}$, and small length difference $b_1-b_2\sim {1\over \sqrt{E}}$.
In periodic orbit theory, the pair correlation function in this limit is called the ``action correlation function"  \cite{argaman1993correlations}.\footnote{In the case of the Riemann zeta function, the analog of the action correlation function is the pairwise distribution of primes given by the Hardy-Littlewood conjecture \cite{bogomolny2003quantum}. }

From an inverse fourier transformation of $K_E(t)$ with respect of $\sqrt{E}$, one can read out the pieces of the pair correlation function that contribute to the ramp-plateau transition.
In particular, for the GUE-like system one gets:
\bea
	f(e^{S_0}y+x,e^{S_0}y-x)=e^{S_0}y\delta(x)+{e^{S_0}\over 2\pi}\int \d k e^{-\i k x}\text{min}(0,2\pi k \rho_0({k^2\over 4})-y)+...\label{eqn:generalf}
\eea
Here we write the expression for general $\rho_0(E)$, and for the Airy density of states it becomes:
\begin{equation}
	f(e^{S_0}y+x,e^{S_0}y-x)\supset{e^{S_0}\over 2\pi}\int dk e^{-\i k x}\text{min}(0,{k^2\over 2}-y)={e^{S_0}\over 2\pi}{x\sqrt{2y} \cos(x\sqrt{2y})-\sin(x\sqrt{2y})\over x^3}. \label{eqn:orbitcorr}
\end{equation}
The $...$ in equation (\ref{eqn:generalf}) represents contributions in the pair correlation that vanish upon Fourier transforming with respect to $x$. 
In particular, since the Airy volumes scales at large genus as:
\be
e^{-2g S_0}V^{\text{Airy}}_g(e^{S_0}y+x,e^{S_0}y-x)\sim {e^{4g S_0} y^{6g}\over \Gamma(4g)},
\ee
 the $...$ contains non-perturbative contributions (in genus expansion parameter) that scale as $e^{e^{S_0}}$.
The $\delta(x)$ piece comes from the cylinder contribution, and the rest presents a finite piece in the pair correlation function from higher genus wormholes that controls the ramp-plateau transition. 
Notice that even though the summation of the volume leads to a large answer for the pair correlation function, only a finite of piece of it determines the ramp-plateau transition. 
In JT gravity, the summation of the volumes is divergent, but the above analysis shows that there is only a finite piece that contributes to the $\tau$-scaled SFF.

In the rest part of this appendix, we will make an observation of the appearance of the formula (\ref{eqn:orbitcorr}) from a direct summation of the Airy volumes. 

Starting from the exact formula of the Airy SFF (\ref{eqn:airySFF}), one can get an exact expression of the Airy volume by stripping off the trumpet wavefunction. 
This leads to:
\be
\begin{split}
	V^{\text{Airy}}_g(b_1,b_2)&=\sum_{n+m=g\,\&\,n,m\geq0}\frac{ (-1)^n b_1^{2 n} b_2^{2 n} 
   (b_1+b_2)^{2 (3 m+n)-2} \Gamma(3m+2n-1/2)}{(2 n+1) 2^{6 m+4 n-1}3^{m}\sqrt{\pi}\Gamma(3m+3n)\Gamma (m+1) \Gamma
   (n+1) \Gamma (3 m+2 n)}\\
   &\times\, _2F_1\left(-3 m-n+1,n+\frac{1}{2};-3m-2 n+3/2;\frac{
   (b_1- b_2)^2}{(b_1+b_2)^2}\right).
\end{split}
\ee
Here the $n$ index comes from the series expansion of the Erf function and the $m$ index comes from the expansion of $\exp(S_0+{1\over 3} e^{-2S_0}\beta^3)$.
Since in the $\tau$-scaled limit, the $e^{-2S_0}\beta^3$ piece is negligible, only the $m=0$ contribution is relevant for the ramp-plateau transition. 
Keeping only the $m=0$ contribution in the sum of $f(e^{S_0}y+x,e^{S_0}y-x)$, one can expand it in power series of $x$ as:
\be 
f(e^{S_0}y+x,e^{S_0}y-x)\supset \sum_{l=0}^{\infty} \mathcal{F}_l(e^{S_0}y^{3/2}) y^lx^{2l}.
\ee 
The $\mathcal{F}_l(e^{S_0}y^{3/2})$ contains the summation of $n$ with $m=0$ and is a complicated generalized hypergeometric function. For instance: 
\begin{align}
	\mathcal{F}_0(e^{S_0}y^{3/2})&=-{1\over 96} e^{4S_0} y^6 \,_2F_5({3\over 4},{5\over 4};1,{4\over 3},{5\over 3},2,{5\over 2};-{1\over 108} e^{4S_0}y^6)\\
	&\approx \left[-{\sqrt{2}\over 3\pi} e^{S_0} y^{3\over 2} + O(e^{-S_0})\right]+ \exp\left({2+2\i \over 3^{3/4}}e^{S_0}y^{3/2}\right)\Bigg[\# y^{-3/2}e^{-S_0} + O(e^{-2S_0})\Bigg].\notag
\end{align}
The large $e^{S_0}$ expansion contains both a ``perturbative'' series and a ``non-perturbative'' series involving $e^{e^{S_0}}$. Note that the non-perturbative piece is numerically much larger, although with a phase that depends rapidly on $S_0$. Interestingly, we found that keeping only the perturbative piece allows us to match with the leading series expansion of the formula in (\ref{eqn:orbitcorr}):
\begin{equation}\label{eqn:sinecosineexpansion}
	{e^{S_0}\over 2\pi}{x\sqrt{2y} \cos(x\sqrt{2y})-\sin(x\sqrt{2y})\over x^3}= -{\sqrt{2}\over 3\pi} e^{S_0} y^{3\over 2}+{\sqrt{2}\over 15\pi} e^{S_0} y^{5\over 2} x^2+...
\end{equation}
We checked this up to order $x^{10}$ by expanding the terms that appear in $\mathcal{F}_l(e^{S_0}y^{3/2})y^l x^{2l}$.

\section{Soft mode action in orbits and gravity}\label{app:soft}
\subsection{Gauge fixing in periodic orbits}
In the main text, encounter configurations were parametrized by the $s$ (stable) and $u$ (unstable) deviations between two orbit segments, measured on a Poincare section at a given time (\ref{srint}). 
\be\label{appendPoin}
\includegraphics[valign = c, scale = .6]{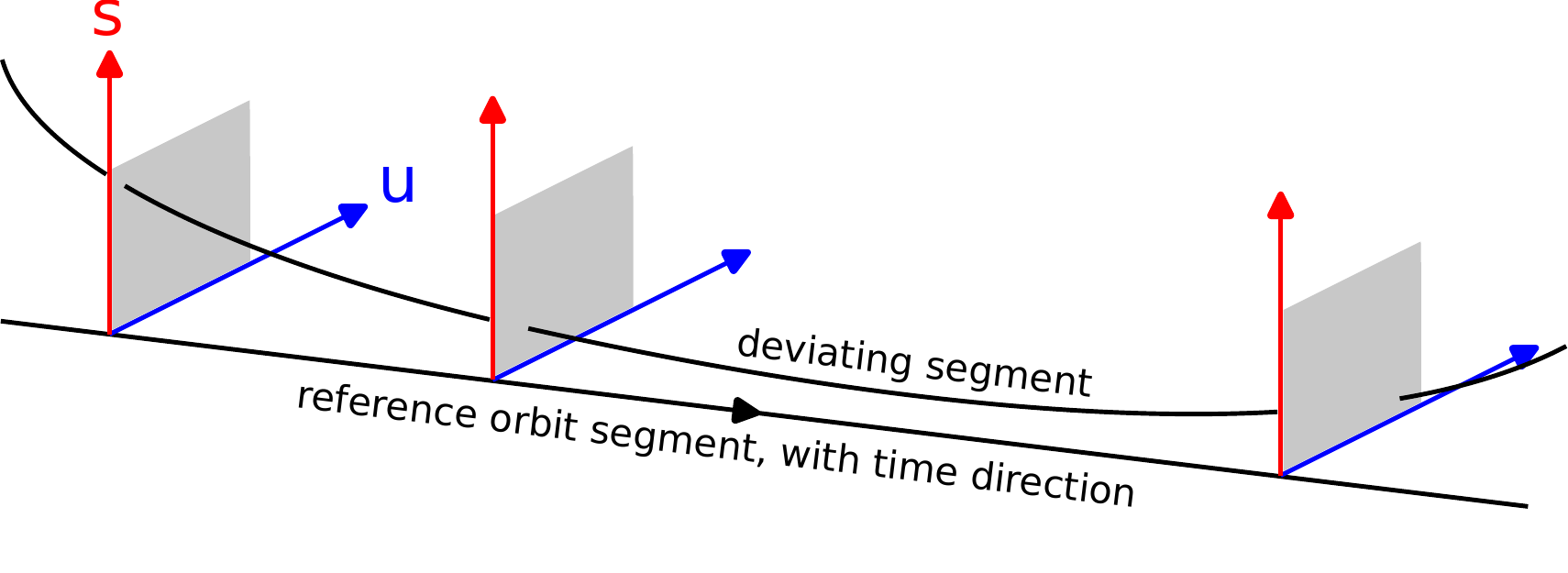}
\ee
For example, in the two-encounter (Sieber-Richter pair), the $s$ and $u$ parameters could measure the deviation of the two portions of the red orbit shown here:
\be
 \centering
    \begin{tikzpicture}[baseline={([yshift=.0ex]current bounding box.center)}, scale=0.7]
 \pgftext{\includegraphics[scale=0.4]{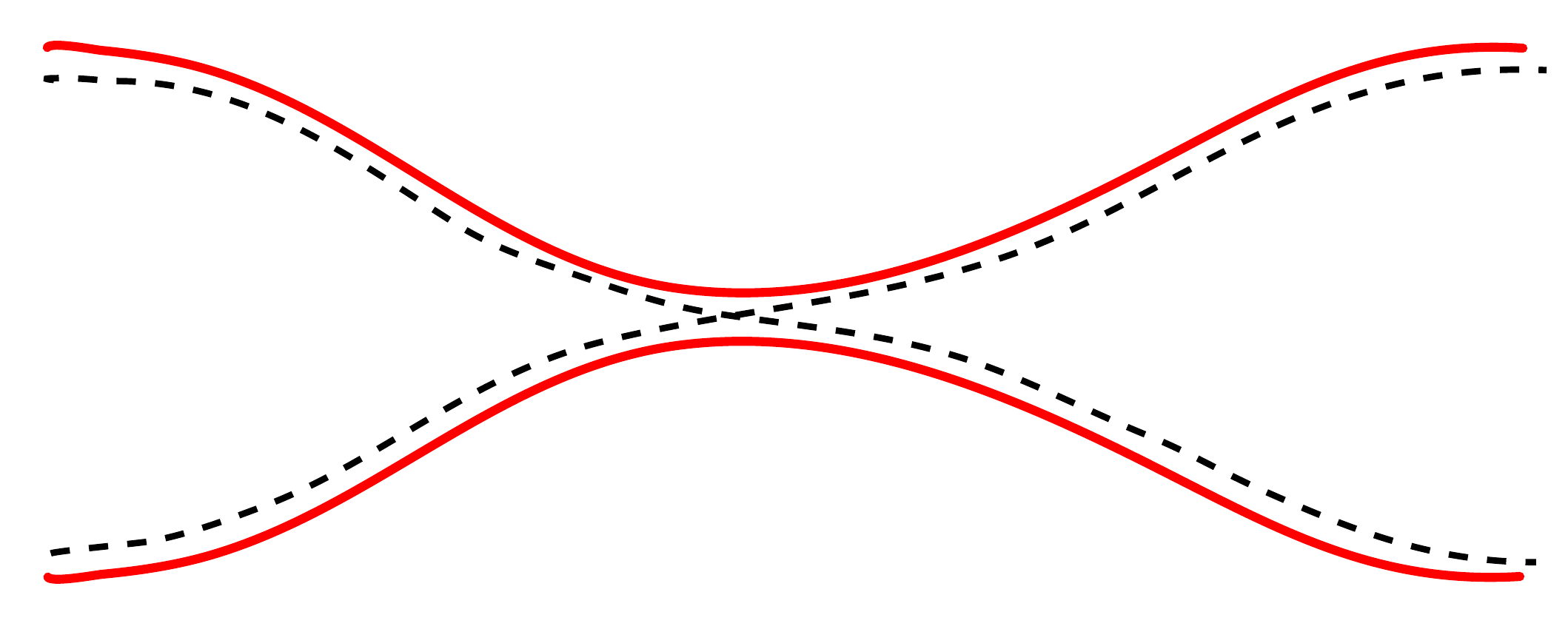}} at (-8,0);

\draw[very thick] (4.1,1.4) circle (0.15);
\draw[very thick] (4.1,-1.4) circle (0.15);

\draw (-4.1, 1.4) node {X};
\draw (-4.1, -1.35) node {X};
  \end{tikzpicture}\quad
\ee
Relative to the red segments, the two black dashed segments have either the $s$ or the $u$ deviation, but not both. In the full Sieber-Richter pair, the pairs of points marked with circles and ``x'' symbols are actually the same point. Each segment runs for time $t/2$, so that when we identify the above points we form orbits of length $t$.

The stable and unstable deviations depend on time, $s(t) \sim e^{-\lambda t}, u(t)\sim e^{\lambda t}$, where $\lambda$ is the Lyapunov exponent, which we will set to one by a rescaling of time. In principle, we can parametrize the encounter by specifying the values of $s \equiv s(t_s)$, $u \equiv u(t_u)$ at different times $t_s,t_u$. This system then has a gauge redundancy under independent shifts of $t_s$ and $t_u$:
\begin{align}
t_s \to t_s  + \alpha_s \hspace{20pt} s \to s e^{-\alpha_s}\\
t_u \to t_u + \alpha_u \hspace{20pt} u \to u e^{\alpha_u}.
\end{align}
In a fully gauge-invariant description of the Sieber-Richter pair, we would integrate over $t_s,t_u,s,u$ and quotient by this gauge group:
\be\label{appendfullint}
K_{E}(t) \supset \frac{t^2}{\pi (2\pi)^2 \rho(E)} \int_{|s(0)|> c,|u(t/2)|> c} \frac{\d t_s \d t_u \d s \d u}{\text{Vol}^2}\exp\left[ (t_s - t_u) + \i e^{t_s - t_u}s u \right].
\ee
The constraints on the $s(0)$ and $u(t/2)$ variables are necessary to make sure that the two segments of the orbit deviate far enough that they can reconnect nontrivially in order to form the Sieber-Richter pair. 

We will consider two different gauge-fixings of this integral:
\begin{enumerate}
\item The periodic orbit integral (\ref{srint}) corresponds to the partial gauge-fixing $t_s=t_u \equiv t_1$. 
\item The JT path integral corresponds to the complete gauge-fixing $t_s=0 $, $t_u=t/2$.
\end{enumerate}
After the first partial gauge-fixing, we find
\be\label{soft1}
K_E(t)\supset \frac{t^2}{\pi (2\pi)^2 \rho(E)} \int_{|s| e^{t_1}> c,|u| e^{\frac{1}{2}t-t_1}> c} \frac{\d t_1\d s \d u }{\text{Vol}}\exp\big[\,\i s u\big].
\ee
In the periodic orbit approach, one only trusts the soft mode action $s u$ for $|s u| < \tilde{c}^2$ for some $\tilde{c}$ that is large in the semiclassical limit. Fortuantely, configurations with larger values of $|su|$ give oscillating contributions that can be ignored in the semiclassical limit. It is convenient to impose the condition $|su| < \tilde c^2$ by writing
\be
\theta(|su| < \tilde c^2) =\int d\gamma  \frac{1}{\tilde t_{enc}}\theta\left(|s| e^{-\gamma}< \tilde c\right) \theta\left( |u| e^{\gamma}< \tilde c \right), \hspace{20pt} \tilde t_{enc} \equiv \ln\frac{\tilde c^2}{|su|}.
\ee
We can insert this into (\ref{soft1}) and then cancel the volume of the gauge group by setting $\gamma = 0$. The result is
\be
\begin{aligned}\label{method1}
K_E(t)&\supset \frac{t^2}{\pi (2\pi)^2 \rho(E)} \int \d t_1\int_{\substack{c e^{-t_1} < |s| < \tilde c \\ c e^{t_1-\frac{1}{2}t} < |u| < \tilde c}}\frac{\d s \d u}{\tilde t_{enc}}\exp\big[\,\i s u\big]
\\&=\frac{t^2}{\pi (2\pi)^2 \rho(E)} \int_{-\tilde c}^{\tilde c} \d s\d u \frac{t-2t_{enc}}{2\tilde t_{enc}}\theta(t > 2 t_{enc})\exp\big[\,\i s u\big].
\end{aligned}
\ee
which is the same as the encounter integral (\ref{srint}). (Here we took care to define separate $c,\tilde{c}$, but this does not affect the result of the integral.)

The second gauge fixing is to pick $t_s=0$, $t_u=\frac{t}{2}$, which will be closely related with the gravity calculation. This leads to
\be\label{appendgauge2}
K_{E}(t) \supset \frac{t^2}{\pi (2\pi)^2 \rho(E)} \int_{|s(0)|> c,|u(t/2)|> c} \d s \d u  \exp\left[-\frac{t}{2} + i  e^{-t/2}su\right].
\ee
The fully gauge-invariant form (\ref{appendfullint}) explains the equivalence between (\ref{method1}) and (\ref{appendgauge2}). Next we will show how the JT results match (\ref{appendgauge2}).

\subsection{Gravity calculation}
It is convenient to start with the formula form the main text (\ref{deltab}) but to undo the integral over the crosscap size parameter $a$:
\be\label{appendCC}
K_{E}(t)\supset e^{-S_0} \frac{t^2\sqrt{E}}{\pi}\int_{-\infty}^{\infty} \d (\delta b) \int_\epsilon^{a^*} \d a e^{i\sqrt{E}\delta b} \frac{1}{\tanh(\frac{a}{4})}.
\ee
In the semiclassical regime of large $E$, the oscillating factor $e^{\i \sqrt{E} \delta b}$ wants to push $\delta b$ into the upper half-plane, and the result of the $a$ integral has a branch point at $\delta b = 2\pi \i$ that dominates the answer. Near the branch point, we rewrite (\ref{CCrelation}) as
\be
\frac{1}{2}e^{\sqrt{E}t}i\sinh\left(\frac{\delta b - 2\pi \i}{4}\right)= \sinh(\frac{a}{4})\sinh(\frac{a'}{4})
\ee
where we also used that $b_1+b_2\approx 4\sqrt{E} t \gg 1$. One can use this to rewrite the integral (\ref{appendCC}) in a more symmetric way by substituting out $a,\delta b$ in favor of 
\be
\hat{X} =\frac{\sinh(\frac{a}{4})}{2^{3/2}E^{1/4}}, \hspace{20pt} \hat{Y} = \frac{\sinh(\frac{a'}{4})}{2^{3/2}E^{1/4}}.
\ee
This leads to
\be
K_{E}(t)\supset -4i  \frac{t^2}{\pi(2\pi)^2 \rho(E)} \int_{C_1} \d\hat{X} \int_{\hat{Y}\neq 0}^{\hat{X}}\d\hat{Y} \exp\left[-\sqrt{E} t + e^{-\sqrt{E}t}\hat{X}\hat{Y}\right]
\ee
The cut off on $\hat{Y}\neq 0$ represents the cutoff $\epsilon$ on the $a$ integral in (\ref{appendCC}). To write the formula in terms of the JT density of states, we used $\rho(E) = e^{S_0}\sinh(2\pi \sqrt{E})/(2\pi)^2 \approx e^{S_0 + 2\pi \sqrt{E}}/8\pi^2$.

The integration contour $C_1$ for $\hat{X}$ is chosen by an analytic continuation of (\ref{CCcutoff}). It goes from $\exp(-i\frac{\pi}{4})\times \infty$ to $\exp(i\frac{\pi}{4})\times \infty$ along the contour below:
\begin{equation}
    \centering
    \begin{tikzpicture}[baseline={([yshift=.0ex]current bounding box.center)}, scale=0.6]
\draw[very thick,->] (-1,0) -- (4.5,0) node[anchor=north west] {$\text{Re}(\hat{X})$};
\draw[very thick,->] (0,-3) -- (0,3) node[anchor=south west] {$\text{Im}(\hat{X})$};
\draw[thick, ->] (0.15,0.15)--(1.4,1.4);
\draw[thick] (1.4,1.4)--(2.8,2.8);

\draw[thick] (1.4,-1.4)--(0.15,-0.15);
\draw[thick,->] (2.8,-2.8)--(1.4,-1.4);
\draw[very thick] (0,0) circle (0.2);

\draw (3.2, 2) node {$\text{ray}\, 1$};
\draw (3.2, -2) node {$\text{ray}\, 2$};

  \end{tikzpicture}\quad
    \label{appendctfig}
\end{equation}

For ray $1$ we do the change of variables $\hat{X}=\exp(i\frac{\pi}{4})X$ and $\hat{Y}=\exp(i\frac{\pi}{4})Y$. For ray $2$, we do change of variables $\hat{X}=\exp(i\frac{3\pi}{4})X$ and $\hat{Y}=\exp(-i\frac{\pi}{4})Y$. In total, we get 
\be\label{appendsoft1}
K_{E}(t)\supset 4  \frac{t^2}{\pi(2\pi)^2 \rho(E)} \int \d X \int_{Y\neq 0}^{|X|}\d Y \exp\left[-\sqrt{E} t + \i e^{-\sqrt{E}t}XY\right]
\ee
This matches the form of (\ref{appendgauge2}) except that we integrate over one quarter of the range, with a factor of four out front. One factor of two arises from the fact that the $|X|>|Y|$ region is equivalent to $|Y|>|X|$ region via mapping class group. The other factor of two is related to the fact that in JT we intgrate over $Y > 0$. It can also been seen from solving a particle in hyperbolic space problem \cite{Sieber_2002}, that $X\to -X$, $Y\to -Y$ together with a change of reference point gives back the same orbit. In JT gravity, we fix such redundancy by only integrate over region $Y>0$.

It would be interesting to understand the analog of the crosscap cylinder contribution in a higher dimensional black hole.

\section{Formula for \texorpdfstring{$P_g^{(\rho)}(\beta)$}{Pg(beta)}}\label{Pappendix}

In this appendix we show that the series expansion for the $\tau$-scaled spectral form factor (\ref{tauexpansionintro}) is
\be\label{kseriesappendix}
e^{-S_0}K_\beta(t) = \frac{\tau}{4\pi \beta} + \sum_{g=1}^\infty P^{(\rho)}_g(\beta)\; \tau^{2g+1},
\ee
with
\be\label{Pdef}
P^{(\rho)}_g(\beta)=-\frac{1}{g(2g+1)(2\pi)^{2g+1}} \oint_0 \frac{\d E}{2\pi i} \frac{e^{-2\beta E}}{\rho_0(E)^{2g}}.
\ee
Without loss of generality, we take the ground state energy $E_0=0$. Note that it is important that only even powers of $1/\rho_0(E)$ appear: the $\sqrt{E}$ branch cut in $\rho_0(E)$ disappears in even powers.

To proceed, we first deform the $E$ contour in (\ref{Pdef}) to a contour $\mathcal{C}$ which surrounds the interval $[0,E_*]$ on the real axis, where $\rho_0(E_*)=\frac{\tau}{2\pi}$. Then plugging (\ref{Pdef}) into (\ref{kseriesappendix}) yields
\begin{align}
e^{-S_0}K_\beta(t)-\frac{\tau}{4\pi\beta}&= - \oint_{\mathcal{C}} \frac{\d E}{2\pi i} e^{-2\beta E} \sum_{g=1}^\infty  \frac{1}{g(2g+1)} \frac{\tau^{2g+1}}{(2\pi)^{2g+1}\rho_0(E)^{2g}}\\
&= - \oint_{\mathcal{C}} \frac{\d E}{2\pi i} e^{-2\beta E} \bigg[\frac{\tau}{\pi}+\Big(\rho_0-\frac{\tau}{2\pi}\Big)\log\Big(1-\frac{\tau}{2\pi \rho_0}\Big)-\Big(\rho_0+\frac{\tau}{2\pi}\Big)\log\Big(1+\frac{\tau}{2\pi \rho_0}\Big)\bigg].\notag
\end{align}
In the first line we moved the sum over $g$ inside the integral. For real $E\in[0, E_*]$ the sum does not converge for any nonzero value of $t$ because $\rho_0$ becomes very small near the origin. However, we can choose the contour $\mathcal{C}$ so that it gives the origin a wide enough berth that $|\rho_0(E)|$ is bounded away from zero everywhere on the contour, and the sum will then converge for small enough $\tau$.

The function in the square brackets has a branch cut along the interval $[0,E_*]$. The contour $\mathcal{C}$ surrounds this interval, so the integral is proportional to the discontinuity across the cut. Using the fact that $\rho_0$ changes sign across the cut, and the discontinuity of the logarithm $\log(-x-i\epsilon) - \log(-x + \i \epsilon)=2\pi i \theta(x)$, we find
\be
\text{Disc}[\text{(Square brackets)}] = 2 \pi i \Big(\rho_0-\frac{\tau}{2\pi}\Big),\hspace{10pt} 0< E < E_*.
\ee
Combining the ramp term with the contribution from the cut, we reproduce (\ref{tauexpansionintro}).
\begin{align}
e^{-S_0}K_\beta(t) &= \frac{\tau}{4\pi \beta} +\int_0^\infty \d E e^{-2\beta E} \Big(\rho_0(E)-\frac{\tau}{2\pi}\Big) \theta(E_*-E)
\cr
&= \int_0^\infty \d E e^{-2\beta E} \text{min}\Big\{\frac{\tau}{2\pi},\rho_0(E)\Big\}.
\end{align}

We are interested in two applications of (\ref{Pdef}). First, we can use it to show that the function (\ref{Pdef}) is indeed proportional to the function $P_g^{(\rho)}(\beta)$ appearing in (\ref{highergenuscutoff}).\footnote{We can absorb the constant of proportionality into the coeffient $C$ in (\ref{highergenuscutoff}).}  To see this, notice that the contour integral in (\ref{Pdef}) picks out the coefficient of $1/E$ in the expansion of $e^{-2\beta E}/\rho_0(E)^{2g}$. The logarithm in (\ref{highergenuscutoff}) comes from the the same term in the expansion of $e^{-2\beta E}/\rho_0(E)^{2g}$ in the integrand of (\ref{highergenuscutoff}), and so is proportional to the same polynomial $P_g^{(\rho)}(\beta)$.

Second, (\ref{Pdef}) can be used to determine the radius of convergence of the genus expansion of the $\tau$-scaled spectral form factor. This reduces to estimating the large $g$ asymptotics of 
\be
\int_0 \frac{\d E}{\rho_0(E)^{2g}}  = \int_0 \frac{\d r}{r^g} \frac{\d E}{\d r}, \hspace{20pt} r(E) \equiv \rho_0^2(E).
\ee
The large $g$ asymptotics of this is determined by the closest singularity of $\d E/dr$ to the origin of the $r$ plane. This corresponds to a location where $\rho'(E) = 0$. In the Airy model there is no such solution, and the radius of convergence is infinite. In the JT model, there is a solution on the negative real $E$ axis, and the radius of convergence is $|\tau| < 2\pi|\rho(E_s)|$ where $\rho'(E_s) = 0$.

\bibliography{references}

\providecommand{\href}[2]{#2}\begingroup\raggedright\begin{thebibliography}{10}

\bibitem{Hawking:1987mz}
S.~W. Hawking, ``{Quantum Coherence Down the Wormhole},''
\href{http://dx.doi.org/10.1016/0370-2693(87)90028-1}{{\em Phys. Lett.}
  {\bfseries B195} (1987) 337}.

\bibitem{Lavrelashvili:1987jg}
G.~V. Lavrelashvili, V.~A. Rubakov, and P.~G. Tinyakov, ``{Disruption of
  Quantum Coherence upon a Change in Spatial Topology in Quantum Gravity},''
  {\em JETP Lett.} {\bfseries 46} (1987) 167--169.
[Pisma Zh. Eksp. Teor. Fiz.46,134(1987)].

\bibitem{Giddings:1987cg}
S.~B. Giddings and A.~Strominger, ``{Axion Induced Topology Change in Quantum
  Gravity and String Theory},''
\href{http://dx.doi.org/10.1016/0550-3213(88)90446-4}{{\em Nucl. Phys.}
  {\bfseries B306} (1988) 890--907}.

\bibitem{Coleman:1988cy}
S.~R. Coleman, ``{Black Holes as Red Herrings: Topological Fluctuations and the
  Loss of Quantum Coherence},''
  \href{http://dx.doi.org/10.1016/0550-3213(88)90110-1}{{\em Nucl. Phys. B}
  {\bfseries 307} (1988) 867--882}.

\bibitem{Maldacena:2004rf}
J.~M. Maldacena and L.~Maoz, ``{Wormholes in AdS},''
  \href{http://dx.doi.org/10.1088/1126-6708/2004/02/053}{{\em JHEP} {\bfseries
  02} (2004) 053}, \href{http://arxiv.org/abs/hep-th/0401024}{{\ttfamily
  arXiv:hep-th/0401024}}.

\bibitem{ArkaniHamed:2007js}
N.~Arkani-Hamed, J.~Orgera, and J.~Polchinski, ``{Euclidean wormholes in string
  theory},'' \href{http://dx.doi.org/10.1088/1126-6708/2007/12/018}{{\em JHEP}
  {\bfseries 12} (2007) 018}, \href{http://arxiv.org/abs/0705.2768}{{\ttfamily
  arXiv:0705.2768 [hep-th]}}.

\bibitem{Saad:2018bqo}
P.~Saad, S.~H. Shenker, and D.~Stanford, ``{A semiclassical ramp in SYK and in
  gravity},''
\href{http://arxiv.org/abs/1806.06840}{{\ttfamily arXiv:1806.06840 [hep-th]}}.

\bibitem{Saad:2019lba}
P.~Saad, S.~H. Shenker, and D.~Stanford, ``{JT gravity as a matrix integral},''
\href{http://arxiv.org/abs/1903.11115}{{\ttfamily arXiv:1903.11115 [hep-th]}}.

\bibitem{Stanford:2019vob}
D.~Stanford and E.~Witten, ``{JT Gravity and the Ensembles of Random Matrix
  Theory},'' \href{http://arxiv.org/abs/1907.03363}{{\ttfamily arXiv:1907.03363
  [hep-th]}}.

\bibitem{Blommaert:2019hjr}
A.~Blommaert, T.~G. Mertens, and H.~Verschelde, ``{Clocks and Rods in
  Jackiw-Teitelboim Quantum Gravity},''
  \href{http://dx.doi.org/10.1007/JHEP09(2019)060}{{\em JHEP} {\bfseries 09}
  (2019) 060}, \href{http://arxiv.org/abs/1902.11194}{{\ttfamily
  arXiv:1902.11194 [hep-th]}}.

\bibitem{Saad:2019pqd}
P.~Saad, ``{Late Time Correlation Functions, Baby Universes, and ETH in JT
  Gravity},''
\href{http://arxiv.org/abs/1910.10311}{{\ttfamily arXiv:1910.10311 [hep-th]}}.

\bibitem{Blommaert:2020seb}
A.~Blommaert, ``{Dissecting the ensemble in JT gravity},''
  \href{http://arxiv.org/abs/2006.13971}{{\ttfamily arXiv:2006.13971
  [hep-th]}}.

\bibitem{Stanford:2021bhl}
D.~Stanford, Z.~Yang, and S.~Yao, ``{Subleading Weingartens},''
  \href{http://dx.doi.org/10.1007/JHEP02(2022)200}{{\em JHEP} {\bfseries 02}
  (2022) 200}, \href{http://arxiv.org/abs/2107.10252}{{\ttfamily
  arXiv:2107.10252 [hep-th]}}.

\bibitem{Almheiri:2019qdq}
A.~Almheiri, T.~Hartman, J.~Maldacena, E.~Shaghoulian, and A.~Tajdini,
  ``{Replica Wormholes and the Entropy of Hawking Radiation},''
  \href{http://dx.doi.org/10.1007/JHEP05(2020)013}{{\em JHEP} {\bfseries 05}
  (2020) 013}, \href{http://arxiv.org/abs/1911.12333}{{\ttfamily
  arXiv:1911.12333 [hep-th]}}.

\bibitem{Penington:2019kki}
G.~Penington, S.~H. Shenker, D.~Stanford, and Z.~Yang, ``{Replica wormholes and
  the black hole interior},'' \href{http://arxiv.org/abs/1911.11977}{{\ttfamily
  arXiv:1911.11977 [hep-th]}}.

\bibitem{Stanford:2020wkf}
D.~Stanford, ``{More quantum noise from wormholes},''
  \href{http://arxiv.org/abs/2008.08570}{{\ttfamily arXiv:2008.08570
  [hep-th]}}.

\bibitem{Belin:2020hea}
A.~Belin and J.~de~Boer, ``{Random Statistics of OPE Coefficients and Euclidean
  Wormholes},'' \href{http://arxiv.org/abs/2006.05499}{{\ttfamily
  arXiv:2006.05499 [hep-th]}}.

\bibitem{Belin:2021ryy}
A.~Belin, J.~de~Boer, and D.~Liska, ``{Non-Gaussianities in the statistical
  distribution of heavy OPE coefficients and wormholes},''
  \href{http://dx.doi.org/10.1007/JHEP06(2022)116}{{\em JHEP} {\bfseries 06}
  (2022) 116}, \href{http://arxiv.org/abs/2110.14649}{{\ttfamily
  arXiv:2110.14649 [hep-th]}}.

\bibitem{Chandra:2022bqq}
J.~Chandra, S.~Collier, T.~Hartman, and A.~Maloney, ``{Semiclassical 3D gravity
  as an average of large-c CFTs},''
  \href{http://arxiv.org/abs/2203.06511}{{\ttfamily arXiv:2203.06511
  [hep-th]}}.

\bibitem{Blommaert:2019wfy}
A.~Blommaert, T.~G. Mertens, and H.~Verschelde, ``{Eigenbranes in
  Jackiw-Teitelboim gravity},''
  \href{http://dx.doi.org/10.1007/JHEP02(2021)168}{{\em JHEP} {\bfseries 02}
  (2021) 168}, \href{http://arxiv.org/abs/1911.11603}{{\ttfamily
  arXiv:1911.11603 [hep-th]}}.

\bibitem{Marolf:2020xie}
D.~Marolf and H.~Maxfield, ``{Transcending the ensemble: baby universes,
  spacetime wormholes, and the order and disorder of black hole information},''
  \href{http://arxiv.org/abs/2002.08950}{{\ttfamily arXiv:2002.08950
  [hep-th]}}.

\bibitem{Okuyama:2020ncd}
K.~Okuyama and K.~Sakai, ``{Multi-boundary correlators in JT gravity},''
  \href{http://dx.doi.org/10.1007/JHEP08(2020)126}{{\em JHEP} {\bfseries 08}
  (2020) 126}, \href{http://arxiv.org/abs/2004.07555}{{\ttfamily
  arXiv:2004.07555 [hep-th]}}.

\bibitem{okounkov2002generating}
A.~Okounkov, ``Generating functions for intersection numbers on moduli spaces
  of curves,'' {\em International Mathematics Research Notices} {\bfseries
  2002} no.~18, (2002) 933--957.

\bibitem{2009arXiv0906.4930A}
A.~{Altland}, P.~{Braun}, F.~{Haake}, S.~{Heusler}, G.~{Knieper}, and
  S.~{M{\"u}ller}, ``{Near action-degenerate periodic-orbit bunches: A skeleton
  of chaos},'' {\em arXiv e-prints} (June, 2009) arXiv:0906.4930,
  \href{http://arxiv.org/abs/0906.4930}{{\ttfamily arXiv:0906.4930 [nlin.CD]}}.

\bibitem{Sieber_2001}
M.~Sieber and K.~Richter, ``Correlations between periodic orbits and their r?le
  in spectral statistics,''
  \href{http://dx.doi.org/10.1238/physica.topical.090a00128}{{\em Physica
  Scripta} {\bfseries T90} no.~1, (2001) 128}.
  \url{https://doi.org/10.1238/physica.topical.090a00128}.

\bibitem{Sieber_2002}
M.~Sieber, ``Leading off-diagonal approximation for the spectral form factor
  for uniformly hyperbolic systems,''
  \href{http://dx.doi.org/10.1088/0305-4470/35/42/104}{{\em Journal of Physics
  A: Mathematical and General} {\bfseries 35} no.~42, (Oct, 2002) L613--L619}.
  \url{https://doi.org/10.1088/0305-4470/35/42/104}.

\bibitem{PhysRevLett.93.014103}
S.~M\"uller, S.~Heusler, P.~Braun, F.~Haake, and A.~Altland, ``Semiclassical
  foundation of universality in quantum chaos,''
  \href{http://dx.doi.org/10.1103/PhysRevLett.93.014103}{{\em Phys. Rev. Lett.}
  {\bfseries 93} (Jul, 2004) 014103}.
  \url{https://link.aps.org/doi/10.1103/PhysRevLett.93.014103}.

\bibitem{PhysRevE.72.046207}
S.~M\"uller, S.~Heusler, P.~Braun, F.~Haake, and A.~Altland, ``Periodic-orbit
  theory of universality in quantum chaos,''
  \href{http://dx.doi.org/10.1103/PhysRevE.72.046207}{{\em Phys. Rev. E}
  {\bfseries 72} (Oct, 2005) 046207}.
  \url{https://link.aps.org/doi/10.1103/PhysRevE.72.046207}.

\bibitem{Okuyama:2021cub}
K.~Okuyama and K.~Sakai, ``{'t Hooft expansion of multi-boundary correlators in
  2D topological gravity},'' \href{http://dx.doi.org/10.1093/ptep/ptab090}{{\em
  PTEP} {\bfseries 2021} no.~8, (2021) 083B03},
  \href{http://arxiv.org/abs/2101.10584}{{\ttfamily arXiv:2101.10584
  [hep-th]}}.

\bibitem{Blommaert:2022lbh}
A.~Blommaert, J.~Kruthoff, and S.~Yao, ``{An integrable road to a perturbative
  plateau},'' \href{http://arxiv.org/abs/2208.13795}{{\ttfamily
  arXiv:2208.13795 [hep-th]}}.

\bibitem{Weber:2022sov}
T.~Weber, F.~Haneder, K.~Richter, and J.~D. Urbina, ``{Constraining
  Weil-Petersson volumes by universal random matrix correlations in
  low-dimensional quantum gravity},''
  \href{http://arxiv.org/abs/2208.13802}{{\ttfamily arXiv:2208.13802
  [hep-th]}}.

\bibitem{Cotler:2016fpe}
J.~S. Cotler, G.~Gur-Ari, M.~Hanada, J.~Polchinski, P.~Saad, S.~H. Shenker,
  D.~Stanford, A.~Streicher, and M.~Tezuka, ``{Black Holes and Random
  Matrices},'' \href{http://dx.doi.org/10.1007/JHEP09(2018)002,
  10.1007/JHEP05(2017)118}{{\em JHEP} {\bfseries 05} (2017) 118},
  \href{http://arxiv.org/abs/1611.04650}{{\ttfamily arXiv:1611.04650
  [hep-th]}}.
[Erratum: JHEP09,002(2018)].

\bibitem{Johnson:2020exp}
C.~V. Johnson, ``{Explorations of nonperturbative Jackiw-Teitelboim gravity and
  supergravity},'' \href{http://dx.doi.org/10.1103/PhysRevD.103.046013}{{\em
  Phys. Rev. D} {\bfseries 103} no.~4, (2021) 046013},
  \href{http://arxiv.org/abs/2006.10959}{{\ttfamily arXiv:2006.10959
  [hep-th]}}.

\bibitem{Eynard:2004mh}
B.~Eynard, ``{Topological expansion for the 1-Hermitian matrix model
  correlation functions},''
  \href{http://dx.doi.org/10.1088/1126-6708/2004/11/031}{{\em JHEP} {\bfseries
  11} (2004) 031}, \href{http://arxiv.org/abs/hep-th/0407261}{{\ttfamily
  arXiv:hep-th/0407261}}.

\bibitem{Eynard:2007kz}
B.~Eynard and N.~Orantin, ``{Invariants of algebraic curves and topological
  expansion},'' \href{http://dx.doi.org/10.4310/CNTP.2007.v1.n2.a4}{{\em
  Commun. Num. Theor. Phys.} {\bfseries 1} (2007) 347--452},
  \href{http://arxiv.org/abs/math-ph/0702045}{{\ttfamily
  arXiv:math-ph/0702045}}.

\bibitem{muller2005periodic}
S.~M{\"u}ller, S.~Heusler, P.~Braun, F.~Haake, and A.~Altland, ``Periodic-orbit
  theory of universality in quantum chaos,'' {\em Physical Review E} {\bfseries
  72} no.~4, (2005) 046207.

\bibitem{berry1985semiclassical}
M.~V. Berry, ``Semiclassical theory of spectral rigidity,'' {\em Proc. R. Soc.
  Lond. A} {\bfseries 400} no.~1819, (1985) 229--251.

\bibitem{heusler2004universal}
S.~Heusler, S.~M{\"u}ller, P.~Braun, and F.~Haake, ``Universal spectral form
  factor for chaotic dynamics,'' {\em Journal of Physics A: Mathematical and
  General} {\bfseries 37} no.~3, (2004) L31.

\bibitem{Gu:2021xaj}
Y.~Gu, A.~Kitaev, and P.~Zhang, ``{A two-way approach to out-of-time-order
  correlators},'' \href{http://dx.doi.org/10.1007/JHEP03(2022)133}{{\em JHEP}
  {\bfseries 03} (2022) 133}, \href{http://arxiv.org/abs/2111.12007}{{\ttfamily
  arXiv:2111.12007 [hep-th]}}.

\bibitem{Altland:2020ccq}
A.~Altland and J.~Sonner, ``{Late time physics of holographic quantum chaos},''
  \href{http://dx.doi.org/10.21468/SciPostPhys.11.2.034}{{\em SciPost Phys.}
  {\bfseries 11} (2021) 034}, \href{http://arxiv.org/abs/2008.02271}{{\ttfamily
  arXiv:2008.02271 [hep-th]}}.

\bibitem{Norbury}
P.~Norbury, ``Lengths of geodesics on non-orientable hyperbolic surfaces,''
  {\em Geometriae Dedicata} {\bfseries 134} no.~1, (2008) 153--176.

\bibitem{Gendulphe}
M.~Gendulphe, ``What's wrong with the growth of simple closed geodesics on
  nonorientable hyperbolic surfaces,'' {\em arXiv preprint arXiv:1706.08798}
  (2017) .

\bibitem{kontsevich1992intersection}
M.~Kontsevich, ``Intersection theory on the moduli space of curves and the
  matrix airy function,'' {\em Communications in Mathematical Physics}
  {\bfseries 147} no.~1, (1992) 1--23.

\bibitem{DoThesis}
N.~Do, ``Intersection theory on moduli space of curves via hyperbolic
  geometry.''. PhD Thesis, The University of Melbourne, 2008.

\bibitem{Whitehead1936equivalent}
J.~H. Whitehead, ``On equivalent sets of elements in a free group,'' {\em
  Annals of mathematics} (1936) 782--800.

\bibitem{Penner1988perturbative}
R.~C. Penner, ``Perturbative series and the moduli space of riemann surfaces,''
  {\em Journal of Differential Geometry} {\bfseries 27} no.~1, (1988) 35--53.

\bibitem{Wegner:1979tu}
F.~Wegner, ``The mobility edge problem: Continuous symmetry and a conjecture,''
  \href{http://dx.doi.org/10.1007/BF01319839}{{\em Zeitschrift f{\"u}r Physik B
  Condensed Matter} {\bfseries 35} no.~3, (1979) 207--210}.
  \url{https://doi.org/10.1007/BF01319839}.

\bibitem{doi:10.1080/00018738300101531}
K.~Efetov, ``Supersymmetry and theory of disordered metals,''
  \href{http://dx.doi.org/10.1080/00018738300101531}{{\em Advances in Physics}
  {\bfseries 32} no.~1, (1983) 53--127},
  \href{http://arxiv.org/abs/https://doi.org/10.1080/00018738300101531}{{\ttfamily
  https://doi.org/10.1080/00018738300101531}}.
  \url{https://doi.org/10.1080/00018738300101531}.

\bibitem{PhysRevLett.75.902}
A.~V. Andreev and B.~L. Altshuler, ``Spectral statistics beyond random matrix
  theory,'' \href{http://dx.doi.org/10.1103/PhysRevLett.75.902}{{\em Phys. Rev.
  Lett.} {\bfseries 75} (Jul, 1995) 902--905}.
  \url{https://link.aps.org/doi/10.1103/PhysRevLett.75.902}.

\bibitem{Berry_1990}
M.~V. Berry and J.~P. Keating, ``A rule for quantizing chaos?,''
  \href{http://dx.doi.org/10.1088/0305-4470/23/21/024}{{\em Journal of Physics
  A: Mathematical and General} {\bfseries 23} no.~21, (Nov, 1990) 4839--4849}.
  \url{https://doi.org/10.1088/0305-4470/23/21/024}.

\bibitem{Keating:1992tq}
J.~Keating, ``The semiclassical functional equation.,''
  \href{http://dx.doi.org/10.1063/1.165919}{{\em Chaos} {\bfseries 2} no.~1,
  (Jan, 1992) 15--17}.

\bibitem{BerryKeatin1992}
M.~V. Berry and J.~P. Keating, ``A new asymptotic representation for zeta
  (1/2+it) and quantum spectral determinants,'' {\em Proceedings: Mathematical
  and Physical Sciences} {\bfseries 437} no.~1899, (1992) 151--173.

\bibitem{argaman1993correlations}
N.~Argaman, F.-M. Dittes, E.~Doron, J.~P. Keating, A.~Y. Kitaev, M.~Sieber, and
  U.~Smilansky, ``Correlations in the actions of periodic orbits derived from
  quantum chaos,'' {\em Physical review letters} {\bfseries 71} no.~26, (1993)
  4326.

\bibitem{Haake:1315494}
F.~Haake, \href{http://dx.doi.org/10.1007/978-3-642-05428-0}{{\em {Quantum
  Signatures of Chaos; 3rd ed.}}}
\newblock Springer series in synergetics. Springer, Dordrecht, 2010.
\newblock \url{https://cds.cern.ch/record/1315494}.

\bibitem{bogomolny2003quantum}
E.~Bogomolny, ``Quantum and arithmetical chaos,'' {\em arXiv preprint
  nlin/0312061} (2003) .

\end{thebibliography}\endgroup


\providecommand{\href}[2]{#2}\begingroup\raggedright\endgroup

\bibliographystyle{utphys}

\end{document}